\documentclass[twocolumn]{aastex62}

\usepackage{times}
\usepackage{amsmath}
\usepackage{graphicx}
\usepackage{hyperref}
\usepackage{gensymb}
\usepackage{upgreek}
\usepackage{natbib}
\usepackage{subfigure}
\usepackage{multirow}
\usepackage{comment}
\usepackage{mathtools}
\usepackage{mathrsfs}
\usepackage{fontenc}
\usepackage{color}
\usepackage{url}
\usepackage{pifont}
\usepackage{ulem}

\newcommand{\Ha}{H$\alpha$}
\newcommand{\Hb}{H$\beta$}
\newcommand{\NeII}{[Ne\,\textsc{ii}]}

\newcommand{\NeV}{[Ne\,\textsc{v}]}
\newcommand{\OII}{[O\,\textsc{ii}]}
\newcommand{\OIII}{[O\,\textsc{iii}]}
\newcommand{\OIV}{[O\,\textsc{iv}]}

\newcommand{\SII}{[S\,\textsc{ii}]}
\newcommand{\NII}{[N\,\textsc{ii}]}

\submitjournal{ApJ}

\shorttitle{Star Formation in Nearby Active Galaxies}
\shortauthors{Zhuang \& Ho}

\begin{document}

\title{The Interplay Between Star Formation and Black Hole Accretion in Nearby Active Galaxies}

\author[0000-0001-5105-2837]{Ming-Yang Zhuang}
\email{mingyangzhuang@pku.edu.cn}
\affil{Kavli Institute for Astronomy and Astrophysics, Peking University,
Beijing 100871, China}
\affil{Department of Astronomy, School of Physics, Peking University,
Beijing 100871, China}

\author[0000-0001-6947-5846]{Luis C. Ho}
\affil{Kavli Institute for Astronomy and Astrophysics, Peking University,
Beijing 100871, China}
\affil{Department of Astronomy, School of Physics, Peking University,
Beijing 100871, China}

\begin{abstract}
Black hole accretion is widely thought to influence star formation in galaxies, but the empirical evidence for a physical correlation between star formation rate (SFR) and the properties of active galactic nuclei (AGNs) remains highly controversial.  We take advantage of a recently developed SFR estimator based on the \OII\ $\lambda3727$ and \OIII\ $\lambda5007$ emission lines to investigate the SFRs of the host galaxies of more than 5,800 type~1 and 7,600 type~2 AGNs with $z < 0.35$.  After matching in luminosity and redshift, we find that type~1 and type~2 AGNs have a similar distribution of internal reddening, which is significant and corresponds to $\sim 10^9\,M_\odot$ of cold molecular gas.  In spite of their comparable gas content, type~2 AGNs, independent of stellar mass, Eddington ratio, redshift or molecular gas mass, exhibit intrinsically stronger star formation activity than type~1 AGNs, in apparent disagreement with the conventional AGN unified model.  We observe a tight, linear relation between AGN luminosity (accretion rate) and SFR, one that becomes more significant toward smaller physical scales, suggesting that the link between the AGN and star formation occurs in the central kpc-scale region.  This, along with a correlation between SFR and Eddington ratio in the regime of super-Eddington accretion, can be interpreted as evidence that star formation is impacted by positive feedback from the AGN.
\end{abstract}


\section{Introduction} \label{sec:intro}

Supermassive black holes (BHs) reside in the centers of all massive galaxies and power their active galactic nuclei (AGNs). The discovery of correlations between the masses of BHs and the properties of their host galaxies \citep{1995ARA&A..33..581Kormendy&Richstone, 1998AJ....115.2285Magorrian+, 2000ApJ...539L...9Ferrarese&Merritt, 2000ApJ...539L..13Gebhardt+} suggests that central BHs coevolve with their host galaxies \citep{1998Natur.395A..14Richstone+, 2004coa..book.....Ho, 2012NewAR..56...93Alexander&Hickox, 2013ARA&A..51..511Kormendy&Ho, 2014ARA&A..52..589Heckman+}.

Numerous observational works have focused on AGNs and their host galaxies, trying to study links between them. However, definitive conclusions remain elusive.  Many report that the strength of AGN activity, usually traced by X-ray luminosity, strongly correlates with the star formation rate (SFR) of their host galaxies \citep[e.g.,][]{2012ApJ...753L..30Mullaney+, 2013ApJ...773....3Chen+, 2015MNRAS.449..373Delvecchio+, 2016MNRAS.457.4179Harris+, 2017A&A...602A.123Lanzuisi+, 2018MNRAS.478.4238Dai+, 2019arXiv191107864Stemo+}. Some find weak or no correlation between the two \citep{2015ApJ...806..187Azadi+, 2015MNRAS.453..591Stanley+, 2017MNRAS.472.2221Stanley+, 2017MNRAS.466.3161Shimizu+}.  Still others maintain that this relation depends on luminosity and redshift: significant correlation between AGN strength and SFR in the most luminous AGNs and at lower redshift, but no correlation for low-luminosity AGNs and sources at higher redshift \citep[e.g.,][]{2010ApJ...712.1287Lutz+, 2012A&A...545A..45Rosario+, 2012A&A...540A.109Santini+}. 

Possible factors that contribute to these conflicting conclusions include small sample size, difficulties of converting photometric measurements to SFRs, selection biases, different binning methods, and the mutual dependence of AGN strength and SFR on stellar mass \citep {2017NatAs...1E.165Harrison}. \citet{2012Natur.485..213Page+} found a negative relation between AGN strength and SFR for X-ray luminosities larger than $10^{44}$ erg~s$^{-1}$. However, \citet{2012ApJ...760L..15Harrison+}, using a larger sample, showed that the results of Page et al. were biased because of sample size. Star-forming galaxies lie on a relation between stellar mass and SFR known as the ``main sequence'' \citep[e.g.,][]{2007ApJ...660L..43Noeske+, 2011A&A...533A.119Elbaz+}. Since the most luminous AGNs tend to reside in the most massive hosts, the mutual dependence of AGN luminosity and SFR on stellar mass could easily lead to the observed correlation between AGN strength and SFR \citep[e.g.,][]{2012A&A...540A.109Santini+, 2017MNRAS.472.2221Stanley+, 2017ApJ...842...72Yang+}.  As discussed in \citet{2014ApJ...782....9Hickox+}, the X-ray or optical continuum traces the instantaneous AGN strength, on timescales much shorter than the timescale for star formation, which is $\ga100$ Myr. In light of the different timescales for the two types of activity and the additional complication introduced by rapid AGN variability, \citet{2014ApJ...782....9Hickox+} predicted that, if SFR and average BH accretion rate are intrinsically coupled, AGN luminosity should be strongly correlated with SFR not at moderate $L_{\rm bol}$ but at high $L_{\rm bol}$.

Estimating SFRs in AGN hosts is nontrivial. The tremendous energy output from accreting BHs can greatly affect or even dominate the entire observed spectral energy distribution, contaminating essentially all SFR indicators conventionally used for star-forming galaxies. Previous works based on X-ray-selected AGN samples predominantly use infrared monochromatic bands or fitting of broad-band spectral energy distributions to calculate the SFR \citep[e.g.,][]{2012ApJ...753L..30Mullaney+, 2012A&A...545A..45Rosario+, 2018MNRAS.478.4238Dai+, 2019arXiv191107864Stemo+}. \citet{2018ApJ...862..118Zhuang+} showed, using a nearby type~1 quasar sample, that if the AGN contribution to the infrared is not properly taken into account, the infrared luminosity from AGN host galaxies can be overestimated by more than three-fold.  Usually only the emission from the AGN torus is considered when estimating the AGN contribution to the infrared. However, the narrow-line region of AGNs also contains dust \citep [e.g.,][]{2003ApJ...587..117Radomski+}, which can be heated by the nucleus and produce strong infrared emission that scales with AGN strength \citep {2006A&A...458..405Groves+, 2011ApJ...732....9Greene+}.   AGN heating can contribute significantly to the far-infrared emission of quasars \citep{2016MNRAS.459..257Symeonidis+}. The strong radiation field of AGNs also likely affects other methods of estimating SFRs, including those that involve ultraviolet and optical photometry \citep{2015ApJ...806..187Azadi+}, emission from polycyclic aromatic hydrocarbons \citep[e.g., ][]{2012ApJ...746..168Diamond-Stanic&Rieke}, and empirically derived emission-line ratios such as \OII\ $\lambda 3727$/\OIII\ $\lambda 5007$ \citep{2005ApJ...629..680Ho, 2009ApJ...696..396Silverman+} and \OIV\ 25.89 \micron/\NeII\ 12.81 \micron\ \citep{2008ApJ...689...95Melendez+, 2010ApJ...725.2270Pereira-Santaella+}. Recent photoionization models based on realistic AGN spectral energy distributions and physical properties of narrow-line regions have led to the development of new methods of estimating SFRs in AGNs using mid-infrared fine-structure neon lines (\NeII\ 12.81 \micron, \NeII\ 15.56 \micron, and \NeV\ 14.32 \micron; \citealt{2019ApJ...873..103Zhuang+, 2019ApJ...875...78Zhuang+}) and optical forbidden oxygen lines (\OII\ $\lambda 3727$ and \OIII\ $\lambda 5007$; \citealt{2019ApJ...882...89Zhuang&Ho}).  

The availability of large optical spectroscopic databases makes the \OII\ and \OIII\ method a powerful tool to study the SFRs of AGN host galaxies.  In this paper, we apply this method to a large sample of more than 5,800 type~1 AGNs and 7,600 type~2 AGNs drawn from the seventh data release \citep[DR7;][]{2009ApJS..182..543Abazajian+} of the Sloan Digital Sky Survey \citep[SDSS;][]{2000AJ....120.1579York+}.  The paper is structured as follows. Section~\ref{sec2} describes the sample selection. In Section~\ref{sec3} we compare the internal galactic extinction of type~1 and type~2 AGNs and study the correlation between AGN properties and the SFRs of their host galaxies.  Section~\ref{sec4} discusses the implications of the results, and our main conclusions are summarized in Section~\ref{sec5}. This paper assumes a cosmology with $H_0=70$ km~s$^{-1}$~Mpc$^{-1}$, $\Omega_m=0.3$, and $\Omega_{\Lambda}=0.7$.  We adopt the stellar initial mass function of \citet{2001MNRAS.322..231Kroupa}.

\section{Sample} \label{sec2}

\subsection{Type~1 AGNs} \label{sec2.1}

\begin{figure*}[t]
\centering
\includegraphics[width=0.49\textwidth]{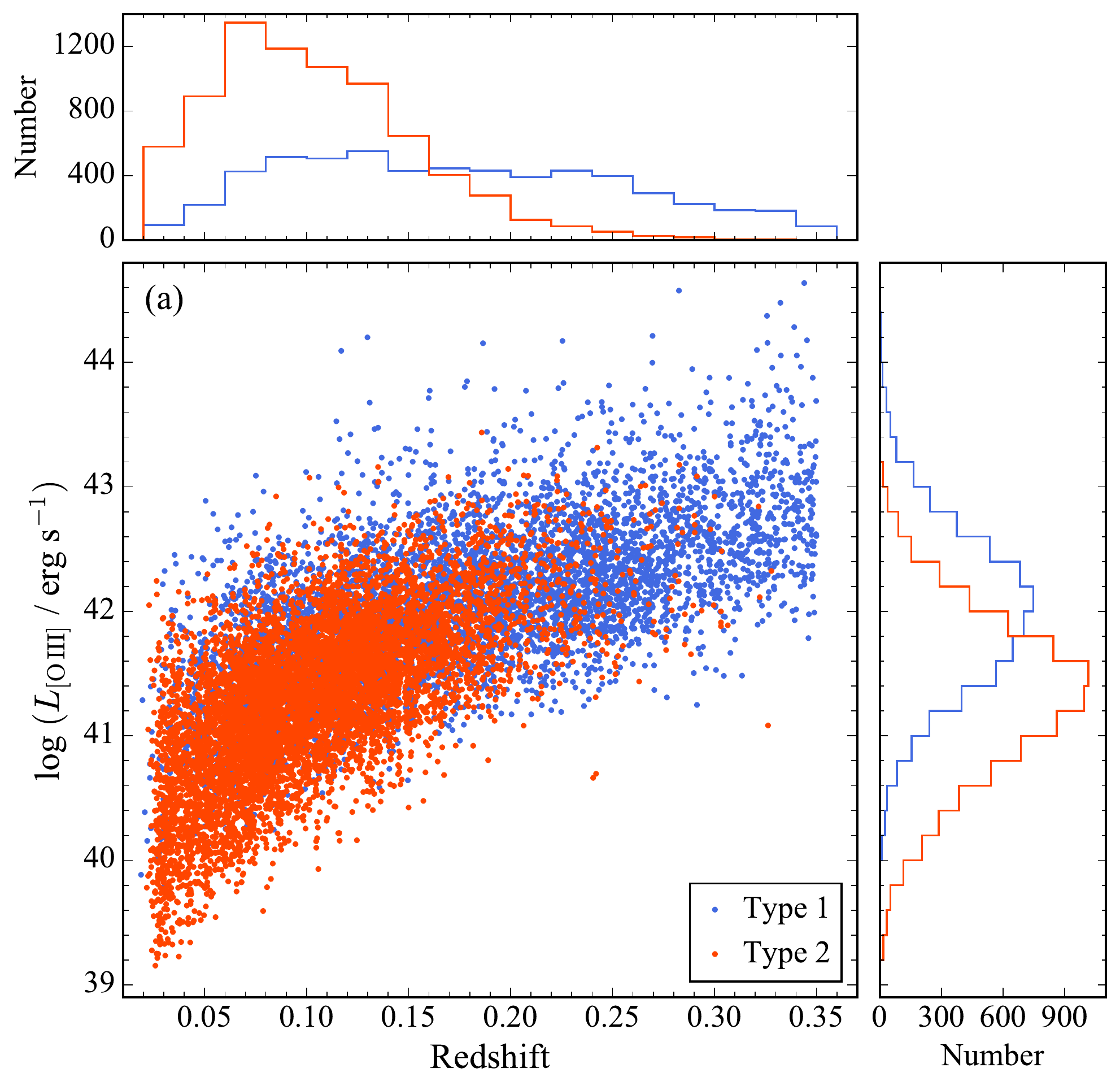}
\includegraphics[width=0.49\textwidth]{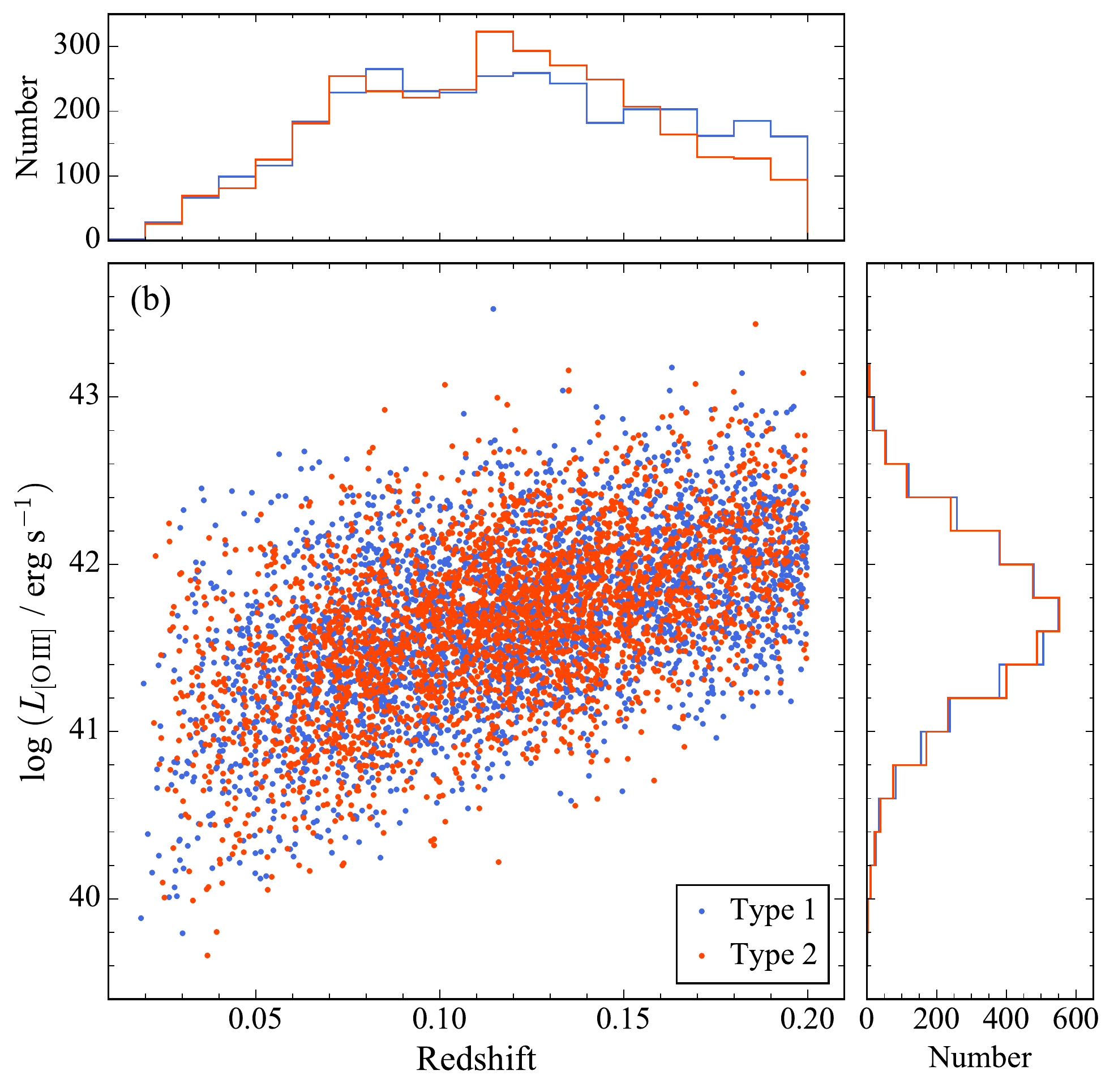}
\caption{Comparison of redshift and extinction-corrected \OIII\ luminosity ($L_{\rm \OIII}$) for (a) parent sample of type~1 (5,838 objects) and type~2 (7,693 objects) AGNs and (b) type~1 and type~2 AGNs after matching their $L_{\rm \OIII}$. Type~1 AGNs are in blue and type~2 AGNs are in red. Top and right panels show the histograms of  redshift and $L_{\rm \OIII}$, respectively.}
\label{fig1}
\end{figure*}

\begin{figure*}[t]
\centering
\includegraphics[width=0.501\textwidth]{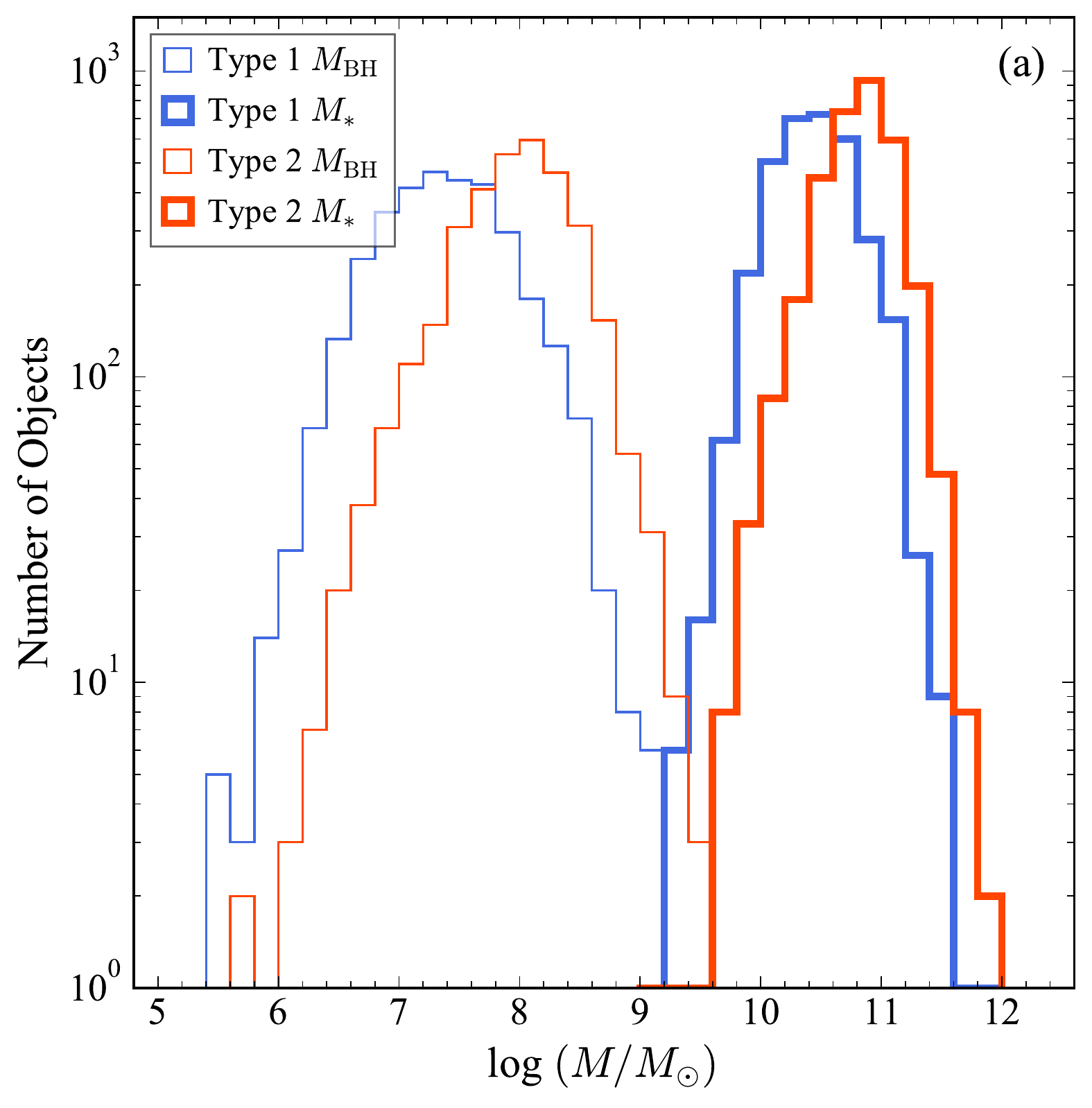}
\includegraphics[width=0.49\textwidth]{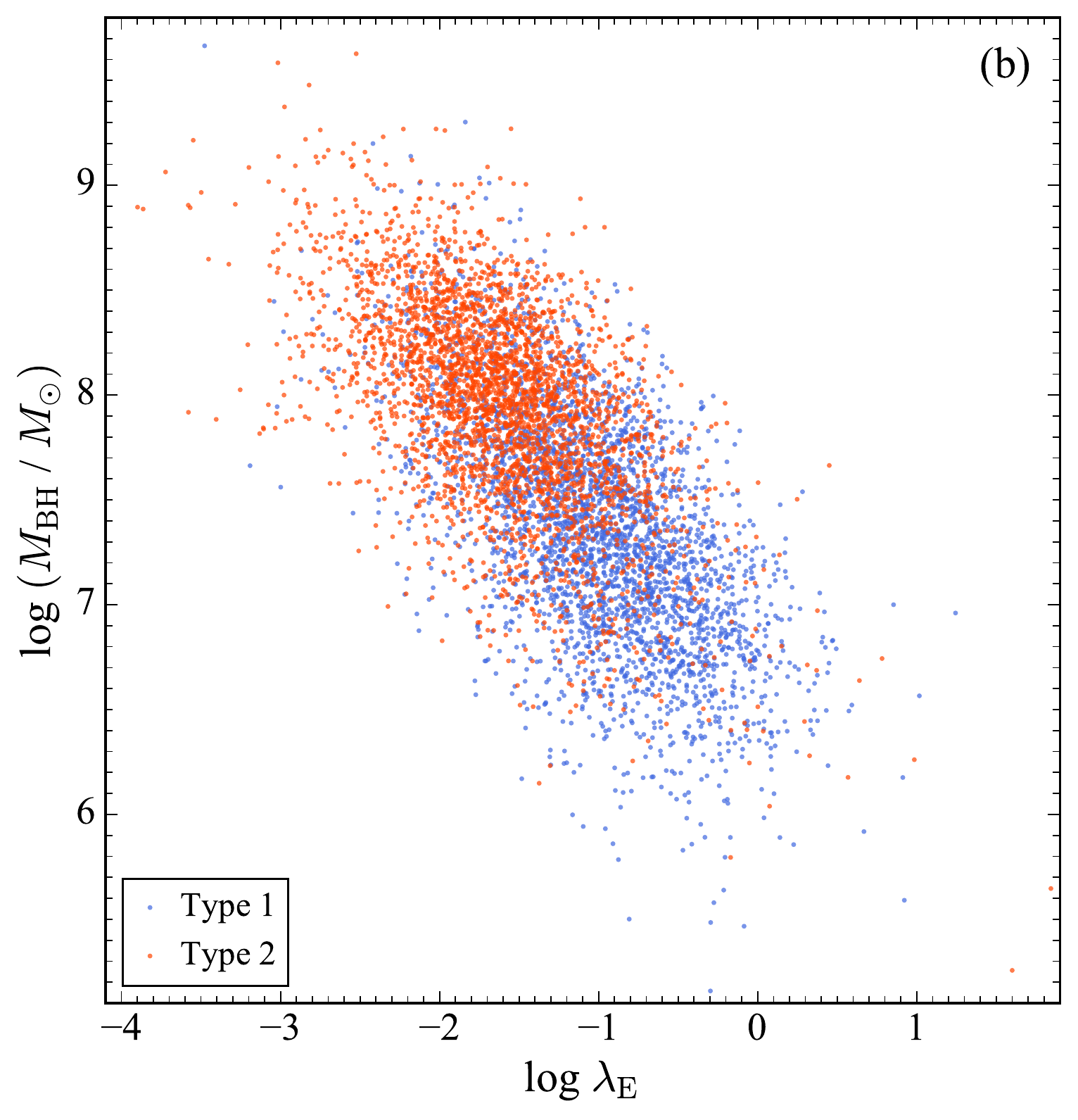}
\caption{(a) Distribution of BH mass ($M_{\rm BH}$; thin lines) and stellar mass ($M_*$; thick lines) for the matched type~1 (blue) and type~2 (red) samples. The values of $M_*$ for type~1s and $M_{\rm BH}$ for type~2s are estimated using Equation~\ref{eq1}. (b) Distribution of $M_{\rm BH}$ versus Eddington ratio ($\lambda_{\rm E}$) for the matched type~1 and type~2 samples.}
\label{fig2}
\end{figure*}

We utilize the sample of 14,584 broad-line (type~1) AGNs with $z<0.35$ from \citet{2019ApJS..243...21Liu+}, which were selected from the database of galaxies and quasars from SDSS DR7. Prior to measuring the emission lines in the spectra, the authors have carefully removed the stellar continuum from the host galaxy and the pseudo-continuum from the AGN. The final sample was determined not based on a fixed value of line width for permitted lines, but on the requirement that broad \Ha\ has larger velocity widths than the narrow lines and several other criteria. We refer readers to \citet{2019ApJS..243...21Liu+} for details. 

Liu et al.'s catalog provides line widths and luminosities for \OIII\ $\lambda 5007$, the narrow and broad components of \Ha\ and \Hb, as well as the AGN continuum luminosity at 5100 \AA. We use their virial BH masses ($M_{\rm BH}$), which are based on the  broad \Ha\ BH mass estimator of \cite{2005ApJ...630..122Greene&Ho}.  In order to obtain the SFRs of AGN host galaxies using the method of \citet{2019ApJ...882...89Zhuang&Ho}, we perform our own measurements of \OII\ $\lambda3727$. We follow the same procedure as \citet{2019ApJS..243...21Liu+} to correct the foreground Galactic extinction and use a single Gaussian plus a linear function to fit the spectra near \OII\ in a small spectral window covering rest wavelengths 3697 \AA\ to 3757 \AA. At the spectral resolution of SDSS ($R\approx 2000$), the \OII\ doublet is barely resolved. We also fit \OIII\ $\lambda5007$ using the same method, and find no systematic difference within 0.08 dex random scatter with respect to the results of \citet {2019ApJS..243...21Liu+}. The narrow emission-line fluxes are corrected for internal dust extinction using the observed Balmer decrement of narrow \Ha\ and \Hb\ and the Milky Way extinction curve of \citet{1989ApJ...345..245Cardelli+} with $R_V=A_V / E(B-V)=3.1$\footnote{Using the extinction curve from \citet{2000ApJ...533..682Calzetti+} with $R_V=4.05$ has only minor effect on the final fluxes for \OII\ (0.03 dex) and \OIII\ (0.05 dex).}. For an electron temperature of $T_e=10^4$ K and electron densities $n_e\approx10^2-10^4$ cm$^{-3}$, the intrinsic value of  \Ha/\Hb\ $= 3.1$ for AGNs \citep{2006agna.book.....Osterbrock&Ferland}. 

We further apply the following selection criteria to the sample:

\begin{enumerate}

\item{A signal-to-noise ratio $\geq$ 3 is required for \OII\ $\lambda3727$, narrow \Hb, \OIII\ $\lambda5007$, narrow \Ha, \NII\ $\lambda6584$, and \SII\ $\lambda\lambda6716, 6731$.} 

\item{We only select objects whose emission lines are dominated by high-excitation AGNs (i.e. Seyferts) based on commonly used emission-line intensity ratios for narrow-line regions \citep{1981PASP...93....5Baldwin+}, focusing on the diagnostic diagrams involving \OIII\ $\lambda5007$/\Hb\ versus \NII\ $\lambda6584$/\Ha\ and \OIII\ $\lambda5007$/\Hb\ versus \SII\ $\lambda\lambda6716, 6731$/\Ha, as defined by the classification boundaries of \citet{2006MNRAS.372..961Kewley+}.  We exclude composite nuclei and low-ionization nuclear emission-line regions \citep[for a review, see][]{2008ARA&A..46..475Ho}, for which the \citet{2019ApJ...882...89Zhuang&Ho} method cannot be applied.}

\item{A signal-to-noise ratio $\geq$ 5 is required for \Ha/\Hb, and a minority of objects with \Ha/\Hb\ $<$ 3.1 are removed, in order to obtain a reliable estimate of extinction.}

\end{enumerate}

Our final sample of type~1 AGNs consists of 5,838 objects, covering 4 orders of magnitude in BH mass ($M_{\rm BH} = 10^{5.4}-10^{9.4}\,M_\odot$) and almost 5 orders of magnitude in AGN luminosity ($L_{\rm \OIII} = 10^{39.8}-10^{44.6}\,{\rm erg~s^{-1}}$).  We derive the bolometric luminosity ($L_{\rm bol}$) and Eddington ratio ($\lambda_{\rm E}\equiv L_{\rm bol}/ L_{\rm E}$) from the extinction-corrected \OIII\ luminosity using a bolometric correction of $L_{\rm bol}= (600\pm150) L_{\rm \OIII}$, as recommended by \citet{2009MNRAS.397..135Kauffmann&Heckman}.  The Eddington luminosity $L_{\rm E} = 1.26\times10^{38}\,(M_{\rm BH}/M_{\odot})\,{\rm erg~s^{-1}}$.  The availability of \OIII\ for both type~1 and type~2 AGNs allows us to compare their properties in a consistent way.  Most previous works used X-ray luminosity to calculate $L_{\rm bol}$.  X-rays trace more instantaneous AGN activity, while \OIII, produced in the extended narrow-line region, captures longer timescales. Using \OIII\ may result in stronger correlation between $L_{\rm bol}$ and SFR.

The stellar masses of the host galaxies of type~1 AGNs are difficult to determine because the nonstellar nucleus strongly contaminates the stellar continuum. Possible ways of estimating the host stellar mass include detailed image decomposition using high-resolution images \citep[e.g.,][]{2009ApJ...701..587Veilleux+, 2017ApJS..232...21Kim+, 2019ApJ...876...35Kim&Ho} and global spectral energy distribution modeling using AGN templates \citep[e.g.,][]{2015A&A...576A..10Ciesla+}.  Recently, \citet{2019arXiv191109678Greene+} presented a scaling relation between $M_{\rm BH}$ and $M_*$ using a large sample of AGNs and quiescent galaxies with diverse morphologies and covering a large dynamical range in $M_{\rm BH}$. Their empirical relation for all galaxy types,

\begin{equation} \label{eq1}
\begin{split}
\log (M_{\rm BH}/M_{\odot}) =  &(7.43 \pm 0.09) + \\
&(1.61\pm0.12) \log(M_*/3\times10^{10}M_{\odot}), 
\end{split}
\end{equation}

\noindent
has significant intrinsic scatter ($0.81\pm0.06$ dex), but it should predict stellar masses without large systematic bias.  Our \OII-based SFR estimator is metallicity-dependent (Equation~\ref{eq2}).  We primarily need host galaxy stellar masses to estimate the metallicity through the mass-metallicity relation (Equation~\ref{eq3}).  Fortunately, the mass-metallicity relation flattens toward high masses and is thus not very sensitive to mass at the high-mass end.  The majority  (85\%) of the objects in our type~1 sample have relatively large stellar masses ($M_* >10^{10.2}\,M_{\odot}$), and hence the approximate stellar masses from Equation~\ref{eq1} suffice.

\subsection{$L_{\rm \OIII}$-matched Type~1 and Type~2 AGN Samples}\label{sec2.2}

For type~2 AGNs, we take the sample from \citet{2019ApJ...882...89Zhuang&Ho}, which is drawn from SDSS DR7 with emission-line fluxes and stellar mass measurements provided by the Max Planck Institute for Astrophysics and Johns Hopkins University (MPA-JHU) catalog\footnote{http://www.strw.leidenuniv.nl/$\sim$jarle/SDSS/, http://www.mpa-garching.mpg.de/SDSS/DR7}. We remove the requirement on fiber specific SFR used in \citet{2019ApJ...882...89Zhuang&Ho}, to match the selection criteria of type~1 sources. The emission-line fluxes are corrected for extinction following the same procedure as for the type~1 sample. The resulting type~2 sample consists of 7,693 objects, slightly larger than the sample of 5,472 objects in \citet{2019ApJ...882...89Zhuang&Ho}.

As the spectroscopic survey of SDSS is magnitude-limited (Petrosian $r<17.77$ mag for the main sample), there are many more type~1 AGNs compared to type~2 AGNs at higher redshifts. We show the distribution of redshift and extinction-corrected $L_{\rm \OIII}$ for the two types of AGNs in Figure~\ref{fig1}a. Type~1 AGNs have a flatter redshift distribution and on average higher $L_{\rm \OIII}$. In order to construct a matched type~2 sample, we set a redshift upper limit of $z=0.2$ and  match their $L_{\rm \OIII}$ distribution using the overlapping region between type~1 and type~2 AGNs using the acceptance-rejection method\footnote{The ratios between the overlapping $L_{\rm \OIII}$ distribution and the $L_{\rm \OIII}$ distribution of the parent samples of type~1 and type~2 AGNs determine the probability of an object to be included into the $L_{\rm \OIII}$-matched sample. For each object, we draw a number from a uniform distribution between 0 and 1. If this number is smaller than the probability for that object, the object is included into the $L_{\rm \OIII}$-matched sample.} \citep{Robert2004}.  As shown in Figure~\ref{fig1}, the resulting distributions of $L_{\rm \OIII}$ for the two types are almost identical, and the redshift distributions are quite similar. 

The matched type~1 and type~2 samples finally contain 3,301 and 3,278 objects, respectively. For consistency with the type~1 sample, we also estimate the BH masses of the type~2 sample using Equation~\ref{eq1}. Both types cover a comparable range of $M_*$ and $M_{\rm BH}$ (Figure~\ref{fig2}a), but the type~2s have systematically somewhat higher fractions of massive BHs and host galaxies. The difference in $M_{\rm BH}$ leads to the difference in $\lambda_ {\rm E}$:  type~2s  comprise most of the population with $\lambda_{\rm E}<0.01$, while the objects with $\lambda_{\rm E} \gtrsim 0.5$  are largely  type~1s  (Figure~\ref{fig2}b).

\section{Results} \label{sec3}

\subsection{Narrow-line Balmer Decrement}\label{sec3.1}

\begin{figure}[t]
\centering
\includegraphics[width=0.5\textwidth]{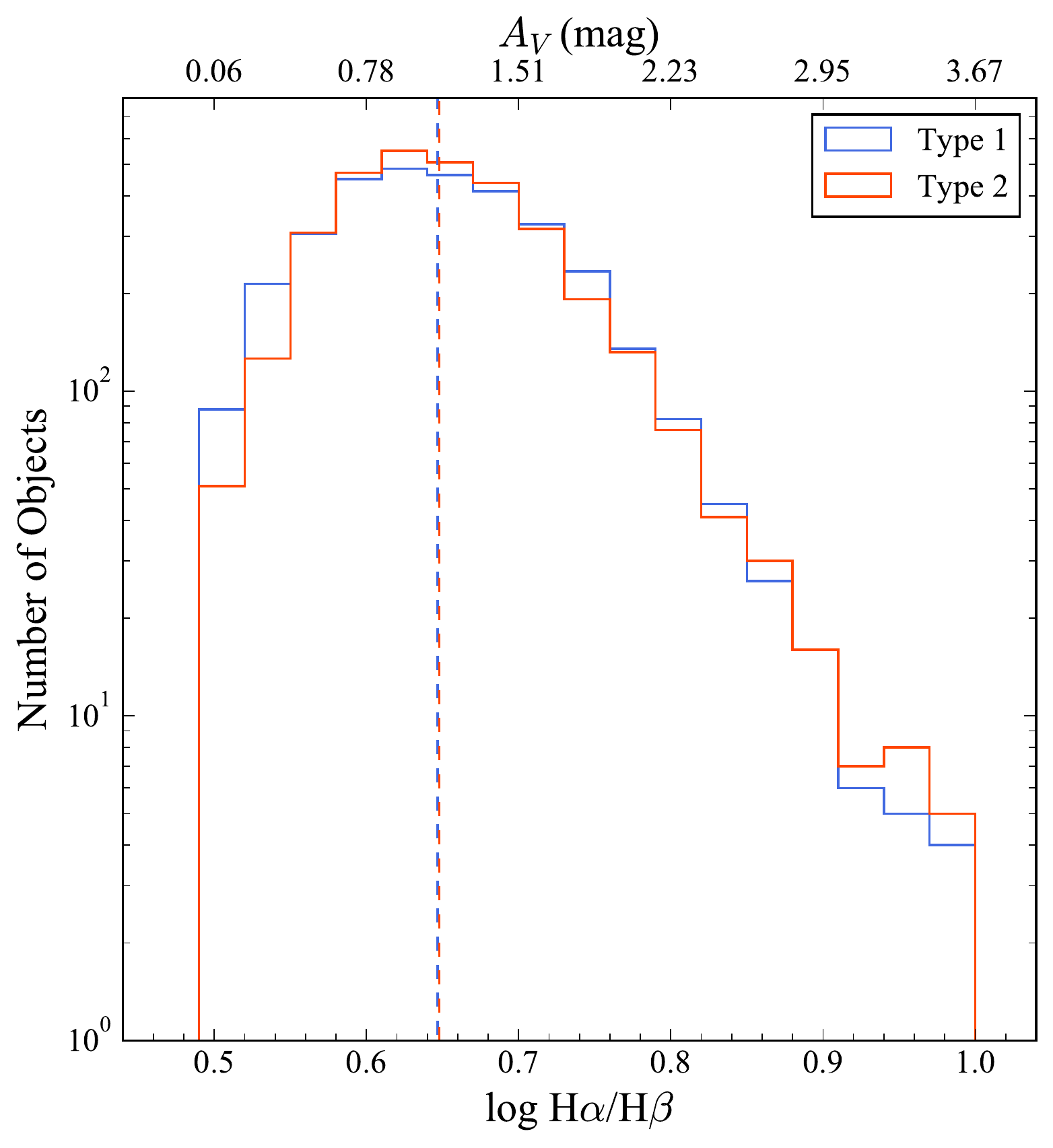}
\caption{Distribution of Balmer decrement (\Ha/\Hb) for the matched samples of type~1 (blue) and type~2 (red) AGNs.  The medians are indicated by dashed vertical lines.  The corresponding extinction ($A_V$) is shown on the top axis.
}
\label{fig3}
\end{figure}

The Balmer decrement of the narrow \Ha\ and \Hb\ lines is widely used as an indicator of internal reddening of the narrow-line regions in type~2 AGNs. Due to difficulties in measuring the narrow component of Balmer lines when the broad component dominates the total flux, the narrow lines of type~1 AGNs are usually not corrected for internal reddening \citep[e.g.,][]{2003MNRAS.346.1055Kauffmann+}. Early work found that type~1 AGNs have a median \Ha/\Hb\ $\approx 3.3$, corresponding to only $A_V \approx 0.2$ mag, consistent with no appreciable extinction in the narrow-line region \citep{2006ApJ...642..702Kim+}. More recent, high-spectral resolution ($R>8000$) observations of narrow hydrogen and helium lines for a small sample of nearby Seyfert 1s suggest that their narrow-line regions experience a larger range of extinctions (up to $A_V \approx 3$ mag; \citealt{2016MNRAS.462.3570Schnorr-Muller+}).  In another study, the distribution of Balmer decrements for a sample of 554 nearby type~1 sources can be well-fitted by a Gaussian with a median \Ha/\Hb\ = 4.37 and a standard deviation of 0.10 dex, further indicating substantial extinction \citep{2019MNRAS.483.1722Lu+}.

How do type~1 and type~2 AGNs compare in terms of their Balmer decrement? In the popular scenario in which gas-rich mergers trigger both AGNs and starbursts \citep[e.g,][]{2005Natur.433..604Di-Matteo+, 2006ApJS..163....1Hopkins+}, type~2 AGNs are obscured by surrounding gas and dust until they grow massive enough and clear the obscuring material to become type~1 systems. For our luminosity-matched type~1 and type~2 AGN samples (Figure~\ref{fig3}), we find that both have relatively similar  Balmer decrement distributions with a median value \Ha/\Hb\ $\approx 4.44$. Our result for type~1 AGNs is in line with \cite{2019MNRAS.483.1722Lu+}. The median $A_V$ estimated using our adopted extinction curve is $\sim1.1$ mag. The close similarity in extinction between the two AGN types indicates that the dust content of their host galaxies is statistically very similar.  As a direct corollary, their gas content must be nearly the same, too, for dust and gas are well-coupled \citep[e.g.,][]{1990ApJ...359...42Devereux&Young}. 

Outflows have been observed in AGNs, both nearby and distant \citep[e.g.,][]{2012ARA&A..50..455Fabian, 2012MNRAS.425L..66Maiolino+, 2018NatAs...2..198Harrison+}. Our results (Figure~\ref{fig3}) suggest that instead of removing gas from their host galaxies, AGN feedback may influence the gas more mildly by disturbing, compressing, or heating it \citep[e.g.,][]{2008ApJ...681..128Ho, 2009ApJ...699..638Ho, 2012NewAR..56...93Alexander&Hickox, 2015ARA&A..53..115King&Pounds, 2017NatAs...1E.165Harrison, 2018MNRAS.478.3447Ellison+, 2018MNRAS.480.3993Baron+, 2018ApJ...854..158Shangguan+, 2019ApJ...870..104Shangguan+, Yesuf2020}. The close similarity between the extinctions of type~1 and type~2 AGNs is also consistent with \citet{2019ApJ...873...90Shangguan&Ho}, who found no differences between the dust masses derived from the 1 to 500 $\mu$m spectral energy distributions of well-matched samples of powerful type~1 and type~2 quasars.   Using the empirical method of \citet{2019ApJ...884..177Yesuf&Ho}, the median extinction of $A_V = 1.1$ mag for our samples of type~1 and type~2 AGNs implies a median molecular gas mass of $10^{8.8}\, M_{\odot}$.  This amount of molecular gas is more or less as expected for nearby, regular star-forming galaxies in this stellar mass range \citep{2017ApJS..233...22Saintonge+}.  Low-redshift AGNs, whether of type~1 or 2, have normal gas content.

\begin{figure*}[t]
\centering
\includegraphics[width=\textwidth]{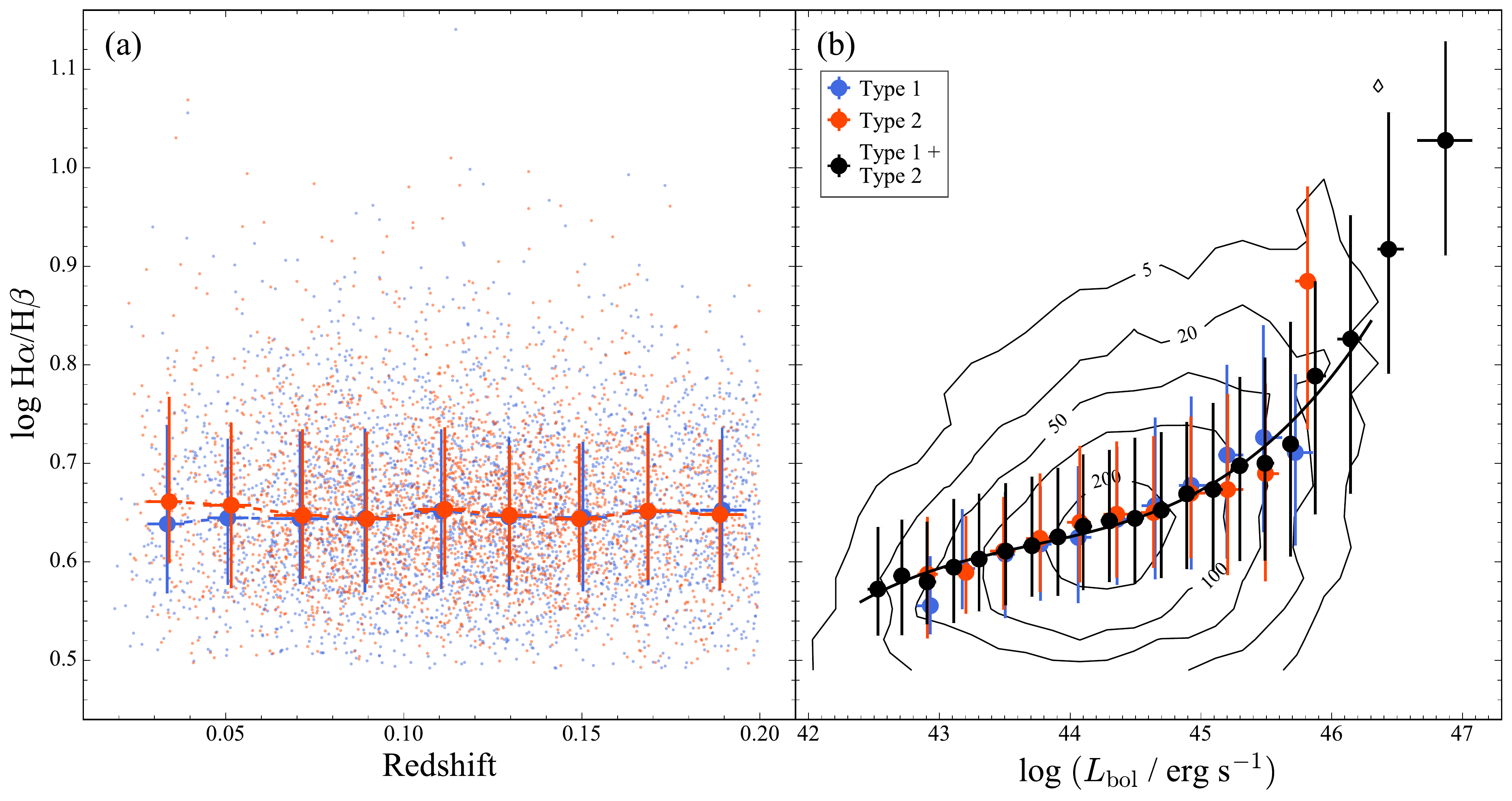}
\caption{Narrow-line Balmer decrement versus (a) redshift and (b) AGN bolometric luminosity. In panel (a), the small blue and red points represent objects from the matched type~1 and type~2 samples, respectively.  The uncertainty of log \Ha/\Hb\ for individual objects is $\sim0.04$ dex; it is not shown here for the sake of clarity. In panel (b), the black contours display the distribution of objects for the combined parent samples of type~1s and type~2s. Large points represent the median of the matched type~1 (blue), type~2 (red), and type~1 + type~2 (black) objects in each bin, with errorbars indicating 16th and 84th percentile.  The solid black curve shows the fit to the median of the parent type~1 + type~2 sample, where the number of objects in each bin exceeds 100.}
\label{fig4}
\end{figure*}

Our objects span a large range of redshift from 0.02 to 0.2, which corresponds to a linear scale of 1.2 kpc to 9.9 kpc for the 3\arcsec-diameter SDSS fiber.  Therefore, it is necessary to investigate whether our result is artificially induced by the increasing physical extent toward higher redshift covered by the fiber. Figure~\ref{fig4}a shows the dependence of \Ha/\Hb\ on redshift for our sample. We find no systematic variation of median \Ha/\Hb\ with redshift, and no difference between type~1s and type~2s. This confirms that our previous result on the distribution of \Ha/\Hb\ (Figure~\ref{fig3}) is not an artifact. On the other hand, we find that \Ha/\Hb\ increases with increasing AGN strength (Figure~\ref{fig4}b), in both the parent sample and luminosity-matched samples of type~1s and 2s. This qualitatively suggests that the large-scale gas content and AGN activity are correlated, and that more luminous AGNs tend to reside in systems with larger gas reservoirs.  For practical purposes, the scaling relation between \Ha/\Hb\ and $L_{\rm bol}$ can be used to estimate the median reddening when \Ha\ shifts outside of the SDSS wavelength coverage for $z \ga 0.35$. A least-squares fit using a third-order polynomial to the median \Ha/\Hb\ in luminosity bins with more than 100 objects gives

\begin{equation} \label{eq4}
\begin{split}
&\log ({\rm H}\alpha/{\rm H}\beta)  =  -(735.993\pm170.880) + (50.4982\pm11.6125)l \\
& - (1.15462\pm0.26298) l^2 + (0.00880498\pm0.00198455)l^3, 
\end{split}
\end{equation}

\noindent 
where $l = \log (L_{\rm bol}/$erg~s$^{-1})$.

\begin{figure*}[t]
\centering
\includegraphics[width=\textwidth]{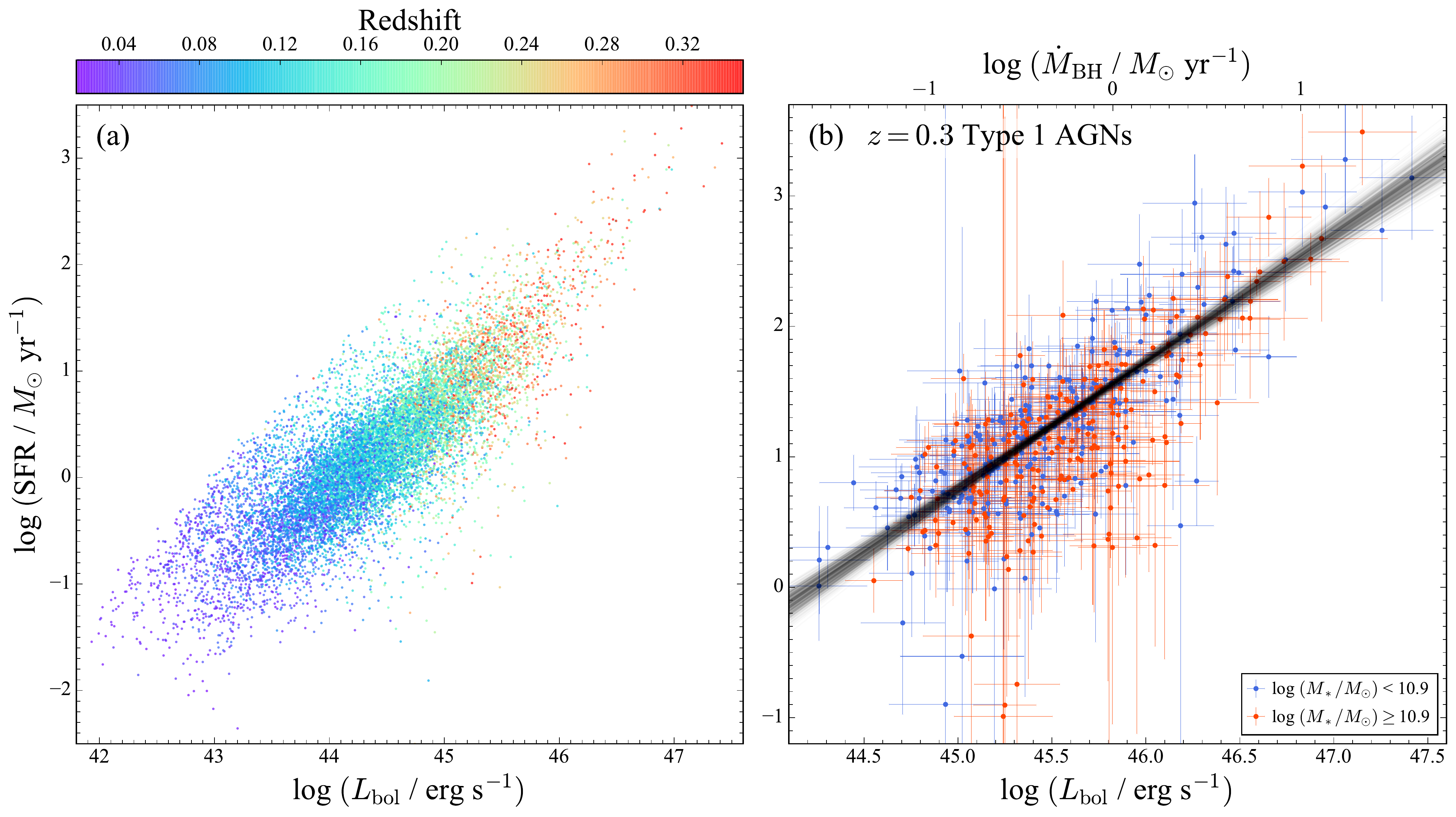}
\caption{(a) Distribution of SFR versus AGN bolometric luminosity ($L_{\rm bol}$) for the parent sample of type~1 and type~2 AGNs, color-coded by redshift spanning 0.02 to 0.35, illustrating the artificial correlation that can be induced by redshift.  (b) The relation between SFR and $L_{\rm bol}$ (equivalently, BH mass accretion rate $\dot{M}_{\rm BH}$; top axis) for the 453 type~1 AGNs at $z=0.3$.  Small dots represent individual objects with errorbars indicating 1~$\sigma$ uncertainty. Objects are color-coded according to their stellar mass, with  log $(M_*/M_{\odot})<10.9$ in blue and $\geq 10.9$ in red.  The black solid line is the best-fit relation for all the objects, and the faint black lines indicate the uncertainty of the fit.}
\label{fig5} 
\end{figure*}

\begin{figure*}[t]
\centering
\includegraphics[width=\textwidth]{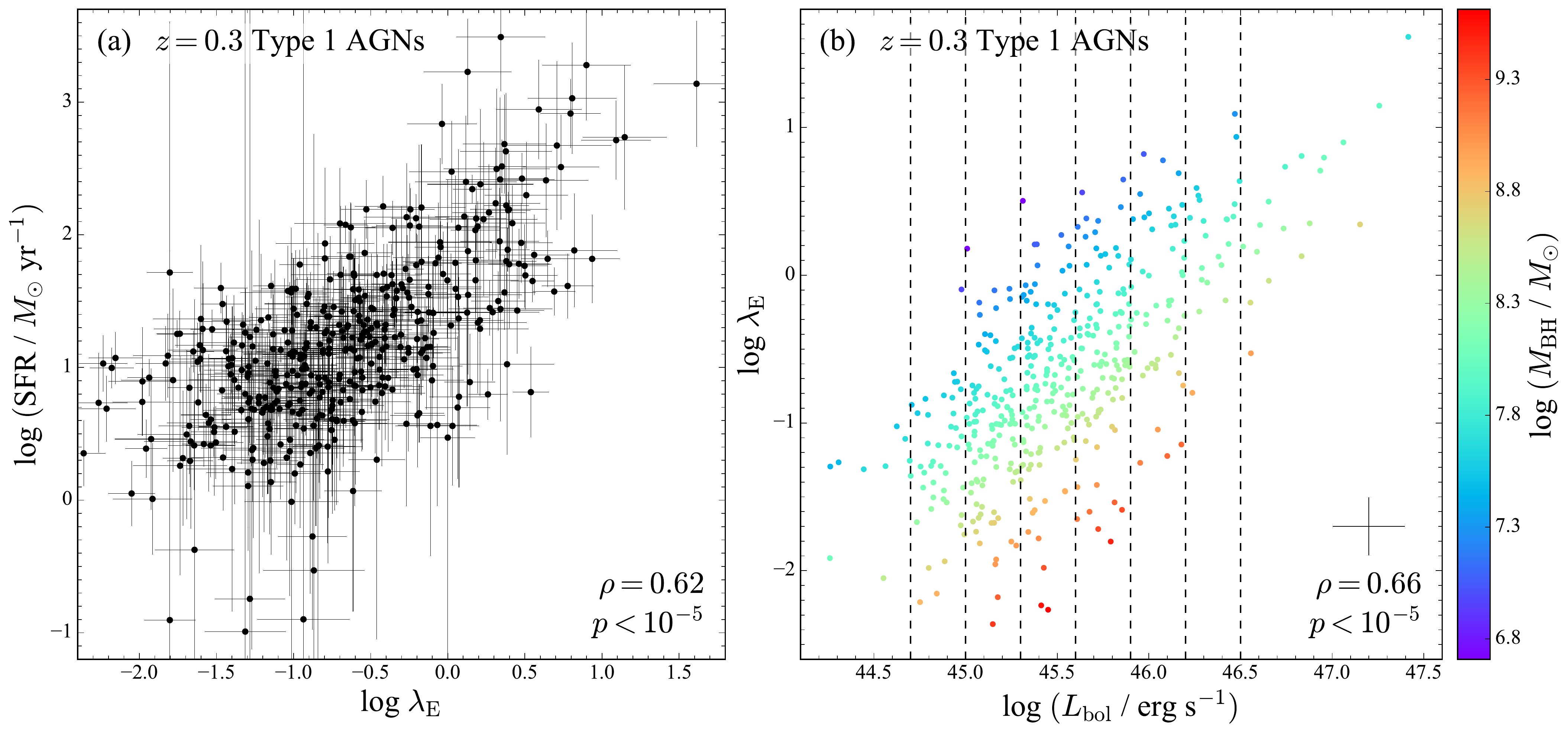}
\caption{
(a) SFR versus $\lambda_{\rm E}$ and (b) $\lambda_{\rm E}$ versus $L_{\rm bol}$ for $z=0.3$ type~1 AGNs, color-coded by $M_{\rm BH}$. The Spearman correlation coefficient and the $p$-value for the whole sample are shown at the lower-right corner of each panel. In panel (b), the errorbars indicate typical uncertainties, and the vertical dashed lines represent the boundaries of $L_{\rm bol}$ subsamples chosen to minimize the dependence of $\lambda_{\rm E}$ on $L_{\rm bol}$.
}
\label{fig6}
\end{figure*}

\begin{figure*}[t]
\centering
\includegraphics[width=\textwidth]{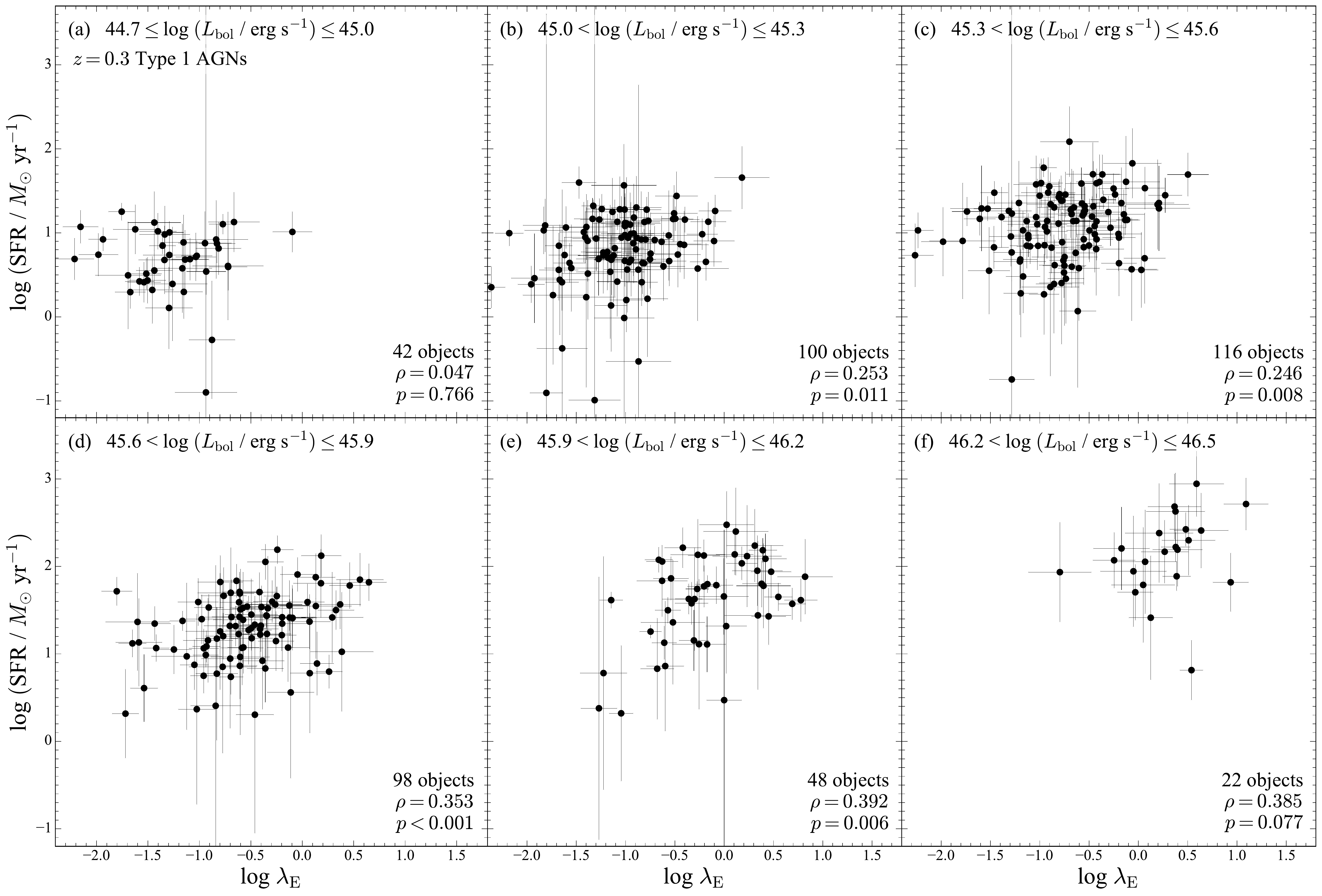}
\caption{
SFR versus $\lambda_{\rm E}$ for six narrow bins in $L_{\rm bol}$ of width 0.3 dex,  for $z=0.3$ type~1 AGNs (see Figure~\ref{fig6}b). The lower-right corner of each panel shows the number of objects in the bin, the Spearman correlation coefficient, and the $p$-value.
}
\label{fig7}
\end{figure*}

\begin{figure*}[t]
\centering
\includegraphics[width=\textwidth]{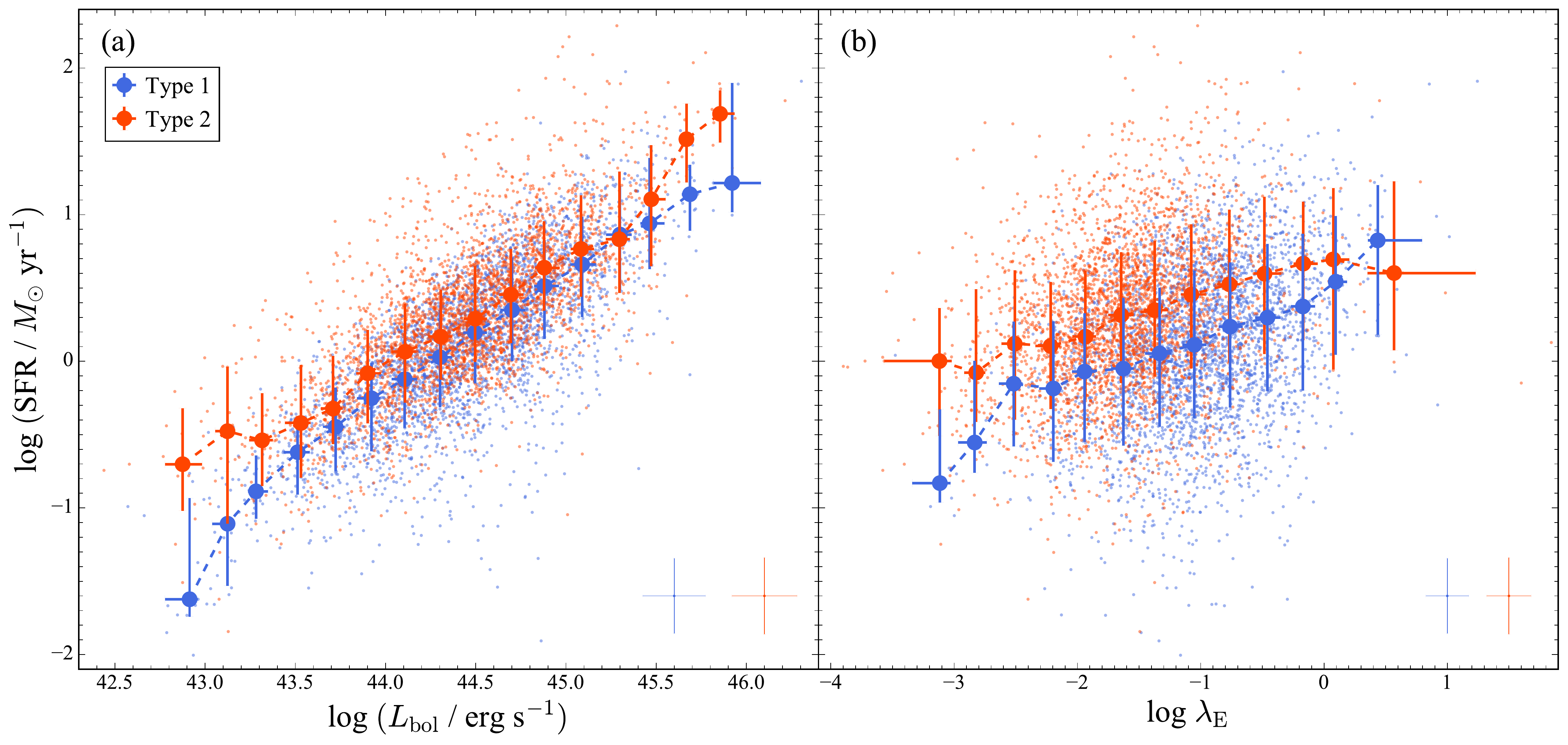}
\caption{Distribution of SFR versus (a) AGN bolometric luminosities ($L_{\rm bol}$) and (b) Eddington ratio ($\lambda_{\rm E}$) for the matched type~1 (blue) and type~2 (red) samples. Small dots represent individual objects, whose median uncertainty is given in the lower-right corner. Large dots represent median values at each $L_{\rm bol}$ and $\lambda_{\rm E}$ bin, with errorbars indicating 16th and 84th percentile.}
\label{fig8}
\end{figure*}

\begin{figure}[t]
\centering
\includegraphics[width=0.5\textwidth]{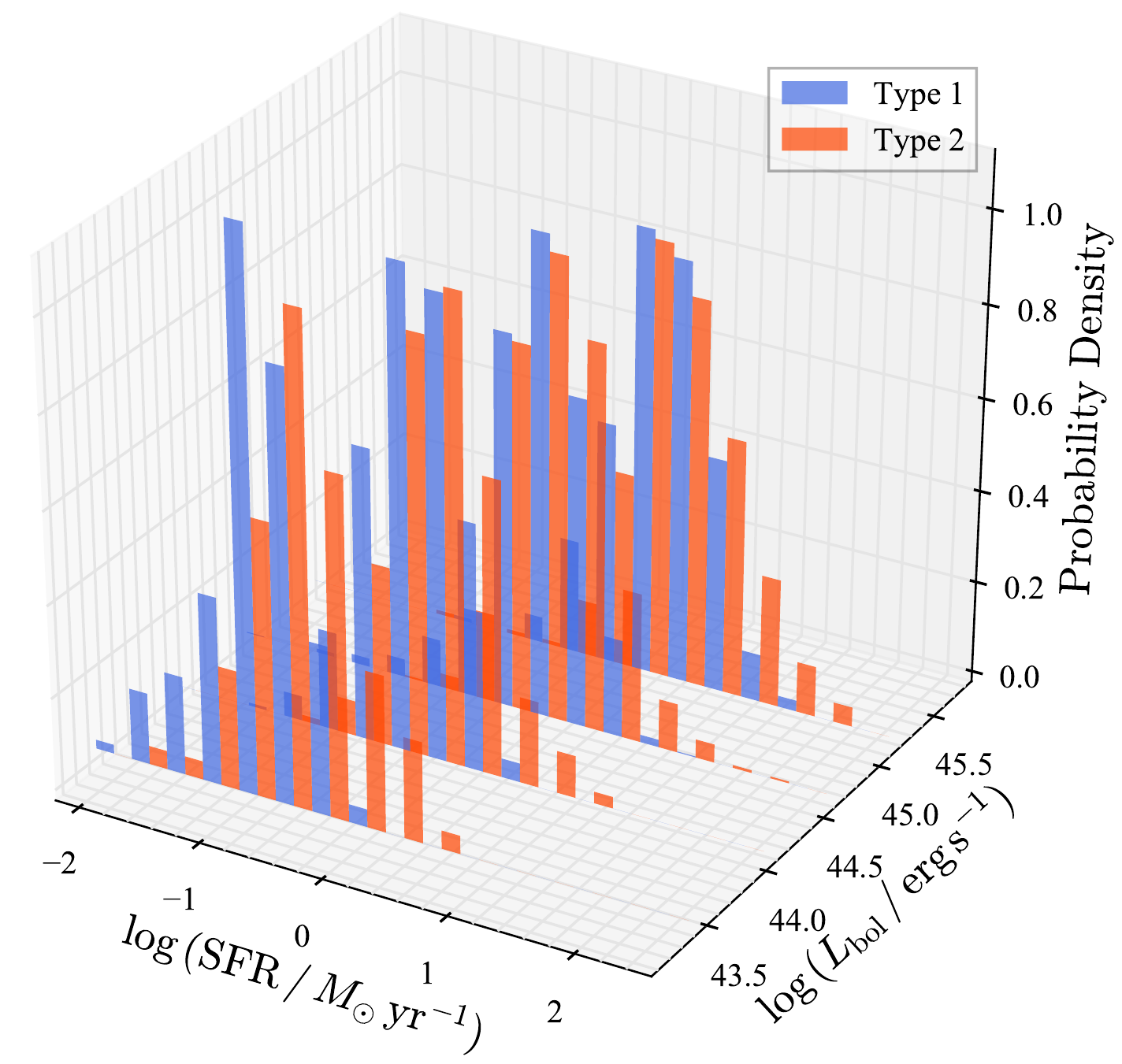}
\caption{Probability density distribution of SFR for the matched type~1 (blue) and type~2 (red) samples, for four bins in $\log (L_{\rm bol}/{\rm erg\,s^{-1}})$: $<43.6$, $43.6-44.3$, $44.3-45$, and $>45$.}
\label{fig9}
\end{figure}

\begin{figure*}[t]
\centering
\includegraphics[width=\textwidth]{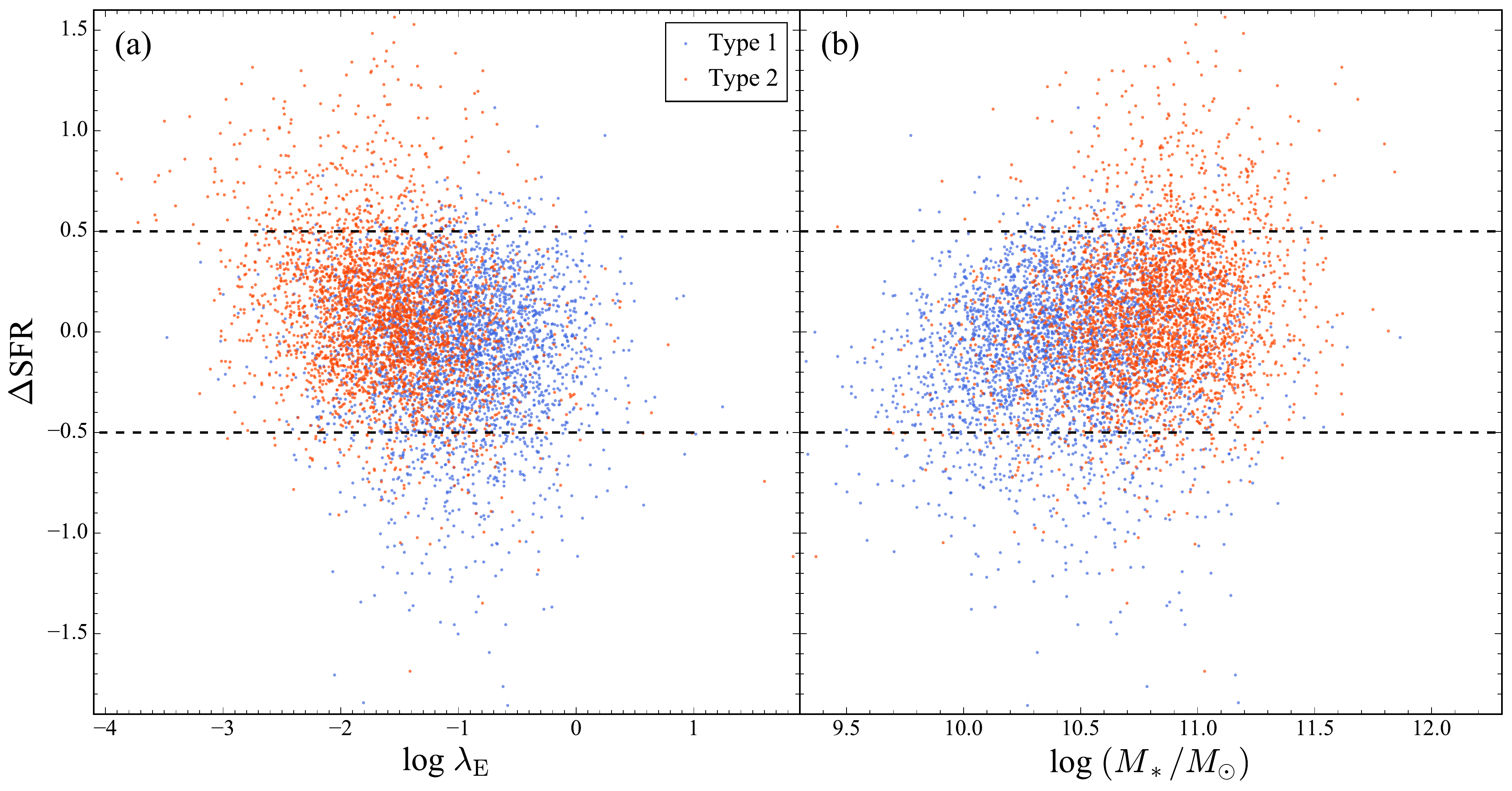}
\includegraphics[width=\textwidth]{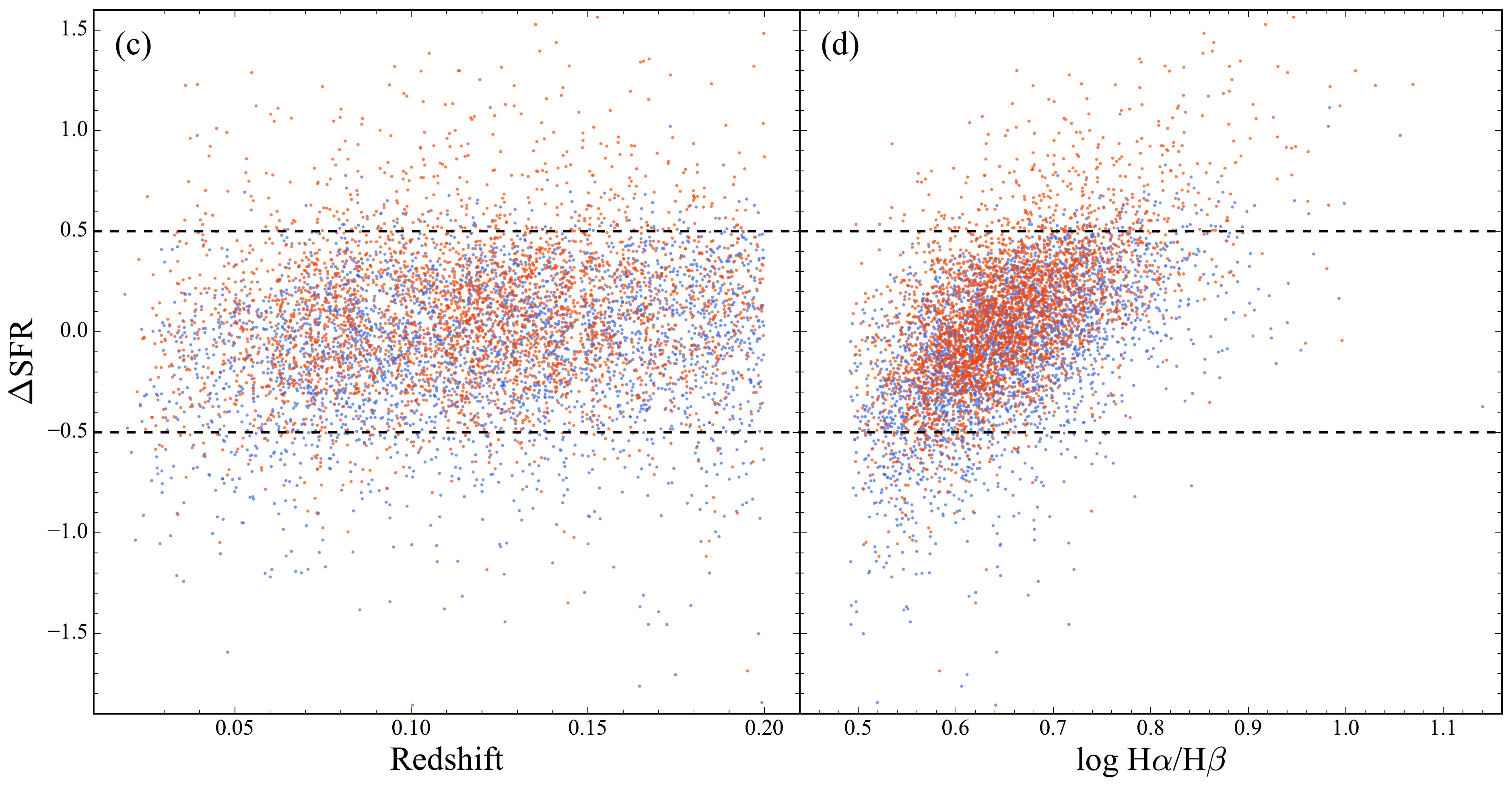}
\caption{Residual SFRs ($\Delta {\rm SFR}\equiv \log {\rm SFR} - \log {\rm SFR}_{\rm median}$) as a function of (a) Eddington ratio, (b) stellar mass, (c) redshift, and (d) Balmer decrement (\Ha/\Hb), where SFR$_{\rm median}$ is the median SFR for matched type~1 (blue) and type~2 (red) samples.  Horizontal black lines ($\pm 0.5$ dex) are plotted to highlight objects with large residual SFRs.}
\label{fig10}
\end{figure*}

\subsection{Estimating SFRs in AGNs}\label{sec3.2}

Motivated by an idea originally suggested by \cite{2005ApJ...629..680Ho}, \citet{2019ApJ...882...89Zhuang&Ho} proposed a new method to estimate the SFRs of AGN host galaxies using the \OII\  $\lambda3727$ and \OIII\ $\lambda5007$ emission lines. Based on photoionization models with realistic AGN intrinsic spectral energy distributions and physical properties of narrow-line regions, \citet{2019ApJ...882...89Zhuang&Ho} constrained the amount of \OII\ emission at a given \OIII\ strength produced by AGNs with relatively high ionization parameters\footnote{AGNs with high ionization parameters, such as Seyferts and quasars, typically have high accretion rates \citep{2009ApJ...699..626Ho}.}. They showed for a large sample of $\sim5,500$ type~2 AGNs that, after accounting for the AGN contribution, SFRs derived from \OII\ are consistent with independent SFR estimates obtained from the stellar continuum of the host galaxies. The proposed \OII\ SFR calibration for AGNs is 

\begin{equation} \label{eq2}
\begin{split}
&{\rm SFR}(M_{\odot}\ {\rm yr}^{-1}) = 5.3\times10^{-42} (L_{\rm \OII}-0.109L_{\rm \OIII}) ({\rm erg\ s^{-1}})/\\
&(-4373.14+1463.92x -163.045x^2+6.04285x^3),
\end{split}
\end{equation}

\noindent
where $L_{\rm \OII}$ and $L_{\rm \OIII}$ are the total extinction-corrected \OII\ and \OIII\ luminosities, and $x =  \log \rm (O/H)+12$ is the oxygen abundance. The contribution to the \OIII\ emission from star formation is negligible because our AGNs predominately resided in massive (Figure~\ref{fig2}a), metal-rich galaxies. As in \citet{2019ApJ...882...89Zhuang&Ho}, we estimate the oxygen abundance from the relation between stellar mass and gas-phase metallicity based on the \NII/\OII\ method of \citet{2008ApJ...681.1183Kewley&Ellison}:

\begin{equation} \label{eq3}
\begin{split}
\rm{log (O/H)+12} = & \, 28.0974 - 7.23631 \log M_* \, + \\ 
                                   & 0.850344 (\log M_*)^2 - 0.0318315 (\log M_*)^3 , 
\end{split}
\end{equation}

\noindent
where the stellar mass $M_*$ is in units of $M_\odot$.  The root-mean-square residual of this relation is 0.10 dex. For objects with stellar masses beyond Kewley \& Ellison's calibration range ($M_* > 10^{11}M_{\odot}$), we fix $\log \rm (O/H)+12$ to 9.02, the value at $M_* = 10^{11}M_{\odot}$. 

We recognize that estimating stellar mass for type~1 AGNs from BH mass using Equation~\ref{eq1} carries significant uncertainty.  However, this does not severely affect our estimation of SFR. As mentioned in Section~\ref{sec2.1}, 85\% of the type~1 AGNs have $M_* >10^{10.2}\,M_{\odot}$. Even in the worst case scenario when the stellar mass of an object with $M_* =10^{10.2}\,M_{\odot}$ is overestimated by 0.8 dex, the overestimate in oxygen abundance by 0.2 dex leads to an underestimate of SFR by only 0.19 dex. This is less than the median uncertainty of 0.29 dex for the SFRs of type 1 AGNs.

\subsection{SFR and BH Accretion Rate}\label{sec3.3}

We investigate the relation between SFR and AGN bolometric luminosities ($L_{\rm bol}$) or, equivalently, BH mass accretion rate, defined as

\begin{equation} \label{eq5}
\dot{M}_{\rm BH}=0.15\left(\frac{\epsilon}{0.1}\right) \left(\frac{L_{\rm bol}}{10^{45}\ {\rm erg\, s^{-1}}}\right)\ M_{\odot}\, {\rm yr}^{-1},
\end{equation}

\noindent
with the radiative efficiency assumed to be $\epsilon = 0.1$.  In order to mitigate against the effect of redshift (Figure~\ref{fig5}a), which induces differences in fiber coverage, we concentrate on the subset of type~1 AGNs lying within the narrow redshift window $z= 0.30-0.35$ (hereinafter the ``$z = 0.3$'' subsample; Figure~\ref{fig5}b). The resulting 453 objects cover  $\log (M_*/M_{\odot}) = 10.0 - 11.8$, with a median of 10.9.  Consistent with previous studies \citep[e.g.,][]{2013ApJ...773....3Chen+, 2016MNRAS.457.4179Harris+, 2018MNRAS.478.4238Dai+}, we find a strong positive correlation between SFR and $\dot{M}_{\rm BH}$ across 3 orders of magnitude in $\dot{M}_{\rm BH}$, with a Spearman correlation coefficient $\rho=0.679$ and $p <10^{-5}$.  A fit using the linear regression code \texttt{linmix} \citep {2007ApJ...665.1489Kelly} gives

\begin{equation} \label{eq6}
\begin{split}
{\rm \log (SFR}/M_{\odot}\ {\rm yr}^{-1}) = & (1.56 \pm 0.02) + \\
& (0.98 \pm 0.04) \log(\dot{M}_{\rm BH}/M_{\odot}\ {\rm yr}^{-1}),
\end{split}
\end{equation}

\noindent
with intrinsic scatter $0.03 \pm 0.01$. The slope is close to unity, indicating that SFR scales linearly with $\dot{M}_{\rm BH}$.  Previous works, which mainly address AGN host galaxies with $\log (M_*/M_{\odot})<11$ \citep[e.g.,][]{2017ApJ...842...72Yang+, 2019arXiv191107864Stemo+}, report a mutual dependence of SFR and $L_{\rm bol}$ ($\propto \dot{M}_{\rm BH}$) on $M_*$.  Our sample does not show this effect.  Splitting our sample into two by the median stellar mass reveals no obvious dependence of Equation \ref{eq6} on $M_*$. Moreover, in the mass range of our sample, the relationship between SFR and $M_*$ is relatively weak because the star-forming ``main sequence'' for local galaxies flattens toward high $M_*$ \citep[e.g.,][]{2019MNRAS.483.3213Popesso+}.

\subsection{SFR versus Eddington Ratio}\label{sec3.4}

SFR scales not only with BH accretion rate, but also with Eddington ratio, albeit with substantial scatter (Figure~\ref{fig6}a).  As in Section~\ref{sec3.3}, we focus on the $z=0.3$ type~1 subsample to avoid the issue of fiber coverage.  For the subsample as whole, we find a highly significant correlation between SFR and $\lambda_{\rm E}$, with $\rho = 0.62$ and $p < 10^{-5}$.  However, this correlation is mostly artificial, driven largely on the one hand by the ${\rm SFR}-L_{\rm bol}$ relation (Figure~\ref{fig5}b), and on the other hand by the strong correlation between $\lambda_{\rm E}$ and $L_{\rm bol}$ owing to the small range of $M_{\rm BH}$ (Figure~\ref{fig6}b).  The $z=0.3$ type~1 AGN sample poorly samples AGNs at low and high $M_{\rm BH}$.

In order to minimize the mutual dependence of SFR and $\lambda_{\rm E}$ on $L_{\rm bol}$, we further divide the sample into six narrow bins of $L_{\rm bol}$, each of width 0.3 dex. For all bins except those of the lowest and highest $L_{\rm bol}$, which suffer from very small-number statistics, a mild correlation between SFR and $\lambda_{\rm E}$ persists (Figure~\ref{fig7}). Moreover, the correlation strength increases systematically with increasing $\lambda_{\rm E}$ (panels b--e). The mass accretion rate onto the BH influences the structure of the accretion flow.  The accretion flow at moderate accretion rates ($0.01\la \lambda_{\rm E} \la 0.1$) is conventionally described by a standard optically thick, geometrically thin disk \citep{1973A&A....24..337Shakura&Sunyaev}, which transitions to an optically and geometrically thick slim disk \citep{1988ApJ...332..646Abramowicz+} at high accretion rates ($\lambda_{\rm E} \ga 0.3$). Our results suggest that SFR and $\lambda_{\rm E}$ tend to be more correlated in the slim disk regime.

\subsection{SFR versus AGN Type}\label{sec3.5} 

The SFRs of the host galaxies of both AGN types strongly correlate with the bolometric luminosities (Figure~\ref{fig8}a).  While the correlation is partly exaggerated by the redshift effect (the matched samples of type~1 and type~2 AGNs span $z \approx 0.02-0.2$), we demonstrated in Section~\ref{sec3.3} that SFR correlates significantly with $L_{\rm bol}$ even after removing the effects of redshift.  Figure~\ref{fig8}b shows that the SFR also correlates with Eddington ratio for both AGN types,  for the overall parent sample as a whole and for each of the types separately after matching in $L_{\rm \OIII}$.  Although differences between two types of AGNs are observed, these trends are difficult to interpret, on account of the effects induced by redshift and the correlation between AGN luminosity and Eddington ratio (Section~\ref{sec3.4}).  Echoing previous studies using SFRs derived from infrared indicators \cite[e.g.,][]{2012A&A...546A..58Rovilos+, 2014MNRAS.437.3550Merloni+, 2019ApJ...878...11Zou+}, there are no significant differences between the median SFRs of the two AGN types.  Instead of looking at all the objects as a group or in a heavily binned manner, the availability of optical spectra for a large number of individual objects enables us to study the detailed distribution of SFRs. The asymmetric errorbars in Figure~\ref{fig8}a clearly reveal that the two AGN types are not evenly distributed on either side of the median $L_{\rm bol}$. For a given bin in $L_{\rm bol}$, type~2 AGNs tend to have moderately more active star formation than type~1 AGNs. In order to better visualize these subtle trends, we further split the sample into four bins in $L_{\rm bol}$ and examine their separate distributions of SFRs (Figure~\ref{fig9}). The distributions of SFR significantly differ for the two AGN types for all the luminosity bins: a two-sample Kolmogorov-Smirnov test yields $p < 10^{-5}$ for the three lower luminosity bins, and  $p \approx 2\times 10^{-3}$  for the highest luminosity bin. 

The two AGN types do not have exactly identical $M_*$, $\lambda_{\rm E}$ (Figure~\ref{fig2}), redshift (Figure~\ref{fig1}b), or Balmer decrement (Figure~\ref{fig3}).  Could this be responsible for the difference in SFR seen here? We investigate the dependence of the difference in SFR distribution on $M_*$, $\lambda_{\rm E}$, redshift, and \Ha/\Hb\ by fitting the median SFRs of the combined type~1 and type~2 samples and calculating the residual SFRs relative to the median ($\Delta$SFR $\equiv$ log~SFR $-$ log~SFR$_{\rm median}$). Interestingly, no obvious trends can be seen between $\Delta$SFR and $\lambda_{\rm E}$, $M_*$, or $z$ (Figure~\ref{fig10}); Type~2 AGNs that scatter by more than 0.5 dex above the median SFR prefer no particular range of these parameters. The same is true for type~1 AGNs whose SFRs deviate by more than 0.5 dex below the median. There is a slight difference for the Balmer decrement, for which we observe a positive trend toward higher \Ha/\Hb. This can be understood as \Ha/\Hb\ is an indicator of gas mass \citep{2019ApJ...884..177Yesuf&Ho}, and objects with larger gas mass tend to have higher SFR.  Still, type~2 AGNs 0.5 dex above the median SFR occupy almost the entire \Ha/\Hb\ range. The above points suggest that, on average, the host galaxies of type~2 AGNs have {\it intrinsically}\ higher SFRs than those of type~1 AGNs, at odds with the conventional unified model of AGNs \citep{1993ARA&A..31..473Antonucci, 2015ARA&A..53..365Netzer}.

\subsection{AGNs Correlate More with Nuclear Star Formation}\label{sec3.6}

\begin{figure}[t]
\centering
\includegraphics[width=0.5\textwidth]{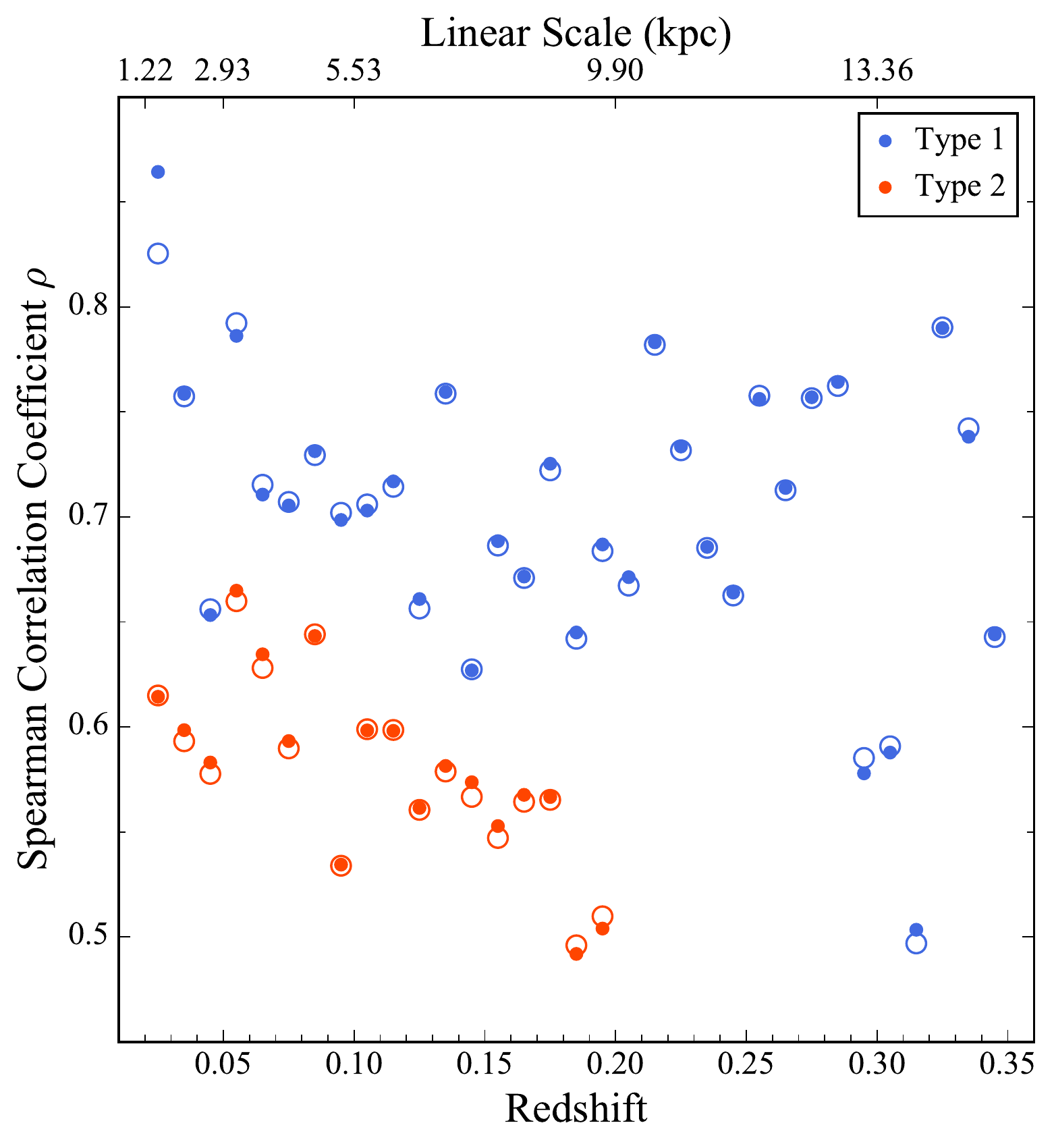}
\caption{Spearman correlation coefficient between SFR and $\dot{M}_{\rm BH}$ as a function of redshift for type~1 (blue) and type~2 (red) samples.  The corresponding linear scale for a 3\arcsec-diameter aperture is given on the top axis. Correlation coefficients based on luminosities are shown in filled circles and those normalized by luminosity distance are shown in open circles. Objects are binned by redshift with bin size 0.01 to avoid redshift coverage difference. We do not include the small number of type~2 AGNs at $z>0.2$.}
\label{fig11}
\end{figure}

In Section~\ref{sec3.3}, we showed that the strong correlation between SFR and $L_{\rm bol}$ is partly, although not wholly, driven by their mutual correlations with redshift (stellar mass). This is because the 3\arcsec-diameter fiber does not cover the entire galaxy at low redshifts (3\arcsec\ corresponds to $\sim 10$ kpc at $z = 0.2$), and the physical area covered by the fiber increases with redshift. Notwithstanding this complication, the fixed angular coverage of the fiber affords an opportunity to investigate the relationship between SFR and AGN strength as a function of physical extent,  down to scales of $\sim1$ kpc. This is of interest in light of previous reports that AGN luminosity connects most intimately with star formation on nuclear scales \citep{2012ApJ...746..168Diamond-Stanic&Rieke, 2014ApJ...780...86Esquej+, 2015MNRAS.449.1470Volonteri+}. To maximize the redshift range and sample size, we use the parent samples of type~1 and type~2 AGNs and bin them in very small (0.01) increments of redshift to ensure similar physical extent within each bin. Figure~\ref{fig11} plots the Spearman correlation coefficient $\rho$ between SFR and $\dot{M}_{\rm BH}$ as a function of redshift (linear scale on top axis). Consistent with previous findings, we also observe a decrease of $\rho$ with increasing physical scale for sources at $z\la 0.2$, both for type~1s and type~2s.  The trend persists even after we normalize SFR and $\dot{M}_{\rm BH}$ by luminosity distance (open symbols). Further dividing the type~2s within each redshift bin into two groups according to $L_{\rm bol}$, we find that more luminous objects tend to have higher $\rho$ than their less luminous counterparts, for virtually all redshift bins. The dependence of $\rho$ on AGN strength explains why the type~1s exhibit systematically higher correlation strengths than the type~2s.  The constancy of $\rho$ beyond $z \approx 0.2$ for type~1 AGNs likely reflects an interplay between the increase of $\rho$ with $L_{\rm bol}$ and the larger physical extent covered by the fiber.  Note that our sample contains too few type~2 AGNs to yield useful results above $z \approx 0.2$.

\section{Implications} \label{sec4}

\subsection{The Impact of AGNs on Star Formation} \label{sec4.1}

Past attempts to study the connection between AGN and star formation activity have yielded highly mixed results, with claimed correlations ranging from strong, to weak, to nonexistent, and still others that are luminosity or redshift-dependent \citep[e.g.,][]{2012A&A...545A..45Rosario+, 2013ApJ...773....3Chen+, 2015ApJ...806..187Azadi+}. It is difficult to fully trace the origin of these inconsistencies, but they arise from a mixture of effects due to sample selection, methods for estimating SFRs and AGN strength, as well as statistical treatment of the data \citep{2015MNRAS.452L...6Volonteri+, 2017NatAs...1E.165Harrison, 2018MNRAS.478.4238Dai+}.  Our study provides a fresh perspective.  Using a large sample of local, optically identified AGNs with robust SFRs estimated for individual galaxies, we demonstrate conclusively that SFR scales tightly and linearly with $L_{\rm bol}$, or, equivalently, $\dot{M}_{\rm BH}$ (Figure \ref{fig5}).  Among the $z \approx 0.3$ type~1 sources used to characterize this result, no obvious differences are found in terms of stellar mass within the range $M_* \approx 10^{10}-10^{12}\, M_{\odot}$.  Our analysis, however, does underscore the necessity of mitigating the effects of redshift and stellar mass, which would otherwise greatly exaggerate the statistical significance of the ${\rm SFR}-L_{\rm bol}$ relation.  These strict requirements stress the vital importance of sample size: without access to the large parent sample enabled by SDSS and our \OII-based SFR estimator, it would be impossible to cull out a subsample of sufficient size and homogeneity to undertake the analysis described in Section~\ref{sec3.3} (Figure~\ref{fig5}b; Equation~\ref{eq6}).

The zero point of the ${\rm SFR}-\dot{M}_{\rm BH}$ relation (Equation~\ref{eq6}) indicates the relative growth of the host galaxy stellar mass and the mass of the central BH, for the population of optically luminous, unobscured AGNs at $z\approx 0.3$. Our sources are highly accreting systems, characterized by $L_{\rm bol} \approx 10^{44.3}-10^{47.4}$~erg~s$^{-1}$, with a median value of $L_{\rm bol} = 10^{45.5}$~erg~s$^{-1}$.  The AGNs in our sample are considerably more powerful than those from previous X-ray-selected samples used for this application \citep[e.g.,][]{2010A&A...518L..26Shao+, 2012A&A...545A..45Rosario+, 2013ApJ...773....3Chen+, 2015ApJ...806..187Azadi+, 2015MNRAS.453..591Stanley+, 2017A&A...602A.123Lanzuisi+, 2017MNRAS.466.3161Shimizu+, 2019arXiv191107864Stemo+}.  Equation~\ref{eq6} gives a zero point $\sim$1.5 dex lower than that of similarly luminous AGNs in higher redshift samples \citep[e.g.,][]{2015MNRAS.449..373Delvecchio+, 2015MNRAS.453..591Stanley+, 2016MNRAS.457.4179Harris+, 2018MNRAS.478.4238Dai+, 2019arXiv191107864Stemo+}.  The increase of average SFR with redshift has also been widely observed in studies using AGN samples covering a large range in redshift \citep[e.g.,][]{2010A&A...518L..26Shao+, 2012A&A...545A..45Rosario+, 2015MNRAS.453..591Stanley+, 2017A&A...602A.123Lanzuisi+}, mirroring the cosmic evolution of the average SFR of star-forming galaxies \citep{2007ApJ...660L..43Noeske+, 2011A&A...533A.119Elbaz+, 2014ARA&A..52..415Madau&Dickinson}.

\subsection{Possible Evidence for Positive AGN Feedback} \label{sec4.2}

What is the physical mechanism that actually causes AGN activity and star formation to go hand-in-hand?  Both processes require fuel, and it is natural to suppose that the gas supply on the very small scales required to feed the BH must be connected, at least loosely, with the gas reservoir on larger scales that forms stars.  We argued (Section~\ref{sec3.6})  that AGN activity couples most closely to stars forming on  central kpc scales. The stronger correlation between $L_{\rm bol}$ and the SFR in the inner parts of the host galaxy suggests a link between AGNs and nuclear star formation.  This qualitatively agrees with the evidence for centrally concentrated star formation in AGN host galaxies gathered through other lines of investigation \citep[e.g.,][] {2013ApJ...765L..33LaMassa+, 2018A&A...609A...9Lutz+, 2019ApJ...873..103Zhuang+}.  If BH accretion relates to star formation, it seems natural to suppose that it would manifest its connection most intimately on nuclear instead of global scales.  Is the connection truly causal or merely incidental?  After all, both BH accretion and star formation share a common gas supply.  A significant fraction of BH growth occurs in starburst galaxies, many of which appear to have been triggered by mergers and galaxy-galaxy interactions \citep{2018MNRAS.480.3201Kauffmann}.  In their recent Atacama Large Millimeter/submillimeter Array study of local bright quasars, \cite{Shangguan2020} note that while the hosts identified with mergers exhibit the highest SFRs, not all hosts with elevated SFRs are associated with mergers.  The authors surmise that under some circumstances AGN feedback may enhance instead of inhibit star formation.

Nearly all of the AGNs in our study have Eddington ratios $\lambda_{\rm E} \gtrsim 0.01$; 31\% lie in the regime of a standard thin accretion disk ($0.01\la \lambda_{\rm E} \la 0.1$), while the rest (33\%) have accretion rates ($\lambda_{\rm E} \ga 0.3$) that formally fall in the territory of a slim disk.  For comparison, most of the X-ray-selected AGN samples in previous studies have significantly lower Eddington ratios \citep[e.g.,][]{2012ApJ...746...90Aird+, 2015ApJ...806..187Azadi+, 2017A&A...601A..63Wang+}.

The strong correlation between SFR and $\lambda_{\rm E}$ (Figure~\ref{fig6}) is largely driven by the mutual dependence of the two variables on $L_{\rm bol}$.  However, a mild correlation still remains after controlling for $L_{\rm bol}$, one that becomes stronger with increasing $\lambda_{\rm E}$ (Figure~\ref{fig7}).  Our results corroborate and place on firmer statistical footing similar trends reported by other investigators.  For example, no significant correlation between SFR and $\lambda_{\rm E}$ was seen in the X-ray-selected AGN sample of \citet{2015ApJ...806..187Azadi+}, which covers X-ray luminosities $\sim 10^{41}-10^{44}\,{\rm erg~s^{-1}}$ and $\lambda_{\rm E} \la 0.1$.  \citet{2016MNRAS.460..902Bernhard+}, also studying AGNs selected through X-rays, reported a slight enhancement of SFR toward higher $\lambda_{\rm E}$, together with a significant increase in starburst fraction. Using polycyclic aromatic hydrocarbon emission as an indicator of recent star formation, \citet{2010MNRAS.403.1246Sani+} showed that narrow-line Seyfert 1 galaxies, usually thought to be high-$\lambda_{\rm E}$ AGNs, experience more intense star formation activity than less highly accreting type~1 Seyferts.  The far-infrared stacking analysis of \citet{2016MNRAS.457.4179Harris+} did not find a strong correlation between $\lambda_{\rm E}$ and SFR for a sample of intermediate-redshift optically selected quasars in the transition range $0.1\la \lambda_{\rm E}<0.6$.  The incidence of AGNs with high specific X-ray luminosities, a proxy for Eddington ratio, is elevated in starburst galaxies \citep{2019MNRAS.484.4360Aird+, 2019MNRAS.487.4071Grimmett+}.

Recent three-dimensional radiation magnetohydrodynamic simulations of the inner regions of accretion flows around supermassive BHs indicate that super-Eddington accretion disks launch high-speed outflows capable of carrying a large fraction of the mass and luminosity \citep{2019ApJ...885..144Jiang+, 2019ApJ...880...67Jiang+}. This theoretical backdrop offers a possible explanation for the observed correlation between SFR and $\lambda_{\rm E}$.  Outflows generated by super-Eddington disks perhaps provide a source of positive feedback on the interstellar medium of the host galaxies.  Compression of the cold molecular gas \citep{2013ApJ...772..112Silk} or direct formation of stars in outflows \citep{2012MNRAS.427.2998Ishibashi&Fabian, 2013MNRAS.431.2350Ishibashi+, 2017Natur.544..202Maiolino+, 2019MNRAS.485.3409Gallagher+} can boost star formation.  Considering that AGNs can vary significantly on timescales of hours to a Myr \citep[e.g.,][]{2011ApJ...737...26Novak+}, outflows from one single epoch of activity are likely delayed with respect to the timescale of AGN activity \citep{2017NatAs...1E.165Harrison}.  Our AGN indicator, \OIII\ $\lambda 5007$ emission, probes AGN activity on relatively long timescales, which, in principle, renders it more sensitive to the cumulative effects of outflows.  As a caveat, we note that our interpretation may be affected by our still relatively limited sample, which is neither large nor complete enough to enable uniform coverage of the high-$\lambda_{\rm E}$ regime at all $L_{\rm bol}$ bins.  In light of the complex and often ambiguous relationship between AGN and star formation tracers \citep{2012A&A...537L...8Cano-Diaz+, 2015ApJ...799...82Cresci+, 2016A&A...591A..28Carniani+}, multi-wavelength observations of the multi-phase properties of outflows of super-Eddington accretors are needed to gain deeper insight into the actual link between BH accretion and galactic-scale star formation.  

\subsection{The Relationship between Type~1 and Type~2 AGNs} \label{sec4.3} 

The unified model of AGNs posits that type~1 and type~2 AGNs are intrinsically the same and differ in their outward manifestation only because of the line-of-sight obscuration by a central dusty torus \citep{1993ARA&A..31..473Antonucci, 2015ARA&A..53..365Netzer}.  The two AGN types should have similar host galaxy properties.  This conventional picture is too idealized.  Galaxies undergo continuous evolution, and so, too, must their nuclear environments.  The rapid evolution of galaxies through gas-rich major mergers provides a natural, alternative physical model for relating the two AGN types.  Within this framework, type~2 AGNs are the precursors of the type~1 counterparts \citep[e.g.,] []{2005Natur.433..604Di-Matteo+, 2006ApJS..163....1Hopkins+}, and naively we do not expect the host galaxies of the two types to be identical.  The evolutionary model for AGNs has received considerable observational support in recent years.  Type~2 quasars possess a number of {\it intrinsic}\ differences relative to type~1 quasars,  including displaying a higher frequency of morphological features consistent with being in an earlier stage of the merger process \citep{2009ApJ...701..587Veilleux+}, higher SFRs \citep{2006ApJ...642..702Kim+, 2008AJ....136.1607Zakamska+, 2016MNRAS.455.4191Zakamska+}, a higher incidence of flat-spectrum radio cores \citep{2010AJ....139.1089Lal&Ho}, and higher Eddington ratios \citep{2018ApJ...859..116Kong&Ho}.  

Our results substantially reinforce these conclusions with a large, statistically significant sample of type~1 and type~2 AGNs matched in redshift and luminosity.  Type~2 AGNs have a consistently larger fraction of elevated SFRs across the range of $L_{\rm bol}$ (Figure~\ref{fig9}), independent of $\lambda_{\rm E}$, $M_*$, or redshift (Figure \ref{fig10}).  Surprisingly, type~1 and type~2 AGNs at the same time have significant but indistinguishable levels of internal dust extinction (Figures~\ref{fig3} and ~\ref{fig10}d), and hence molecular gas content \citep{2019ApJ...884..177Yesuf&Ho}.  \cite{2019ApJ...873...90Shangguan&Ho} arrived at the same conclusion in their far-infrared study of dust masses of matched samples of type~1 and type~2 quasars.  Taken at face value, the combination of higher SFRs but similar molecular gas masses implies that on average type~2 AGNs have higher star formation efficiencies than type~1 AGNs.  This introduces yet another clue into the growing set of physical properties that distinguishes between the two AGN types.

\section{Summary} \label{sec5}

We use more than 5,800 type~1 and 7,600 type~2 low-redshift AGNs to study their star formation activity using a newly developed SFR estimator based on the \OII\ $\lambda3727$ and \OIII\ $\lambda5007$ emission lines.  Careful construction of $L_{\rm \OIII}$-matched subsamples of the two AGN types allows us to deduce a number of conclusions concerning the connection between star formation and the properties of AGNs and their host galaxies:

\begin{enumerate}

\item{Type~1 and type~2 AGNs have a similar distribution of internal extinction, indicating that their host galaxies share very similar dust and gas content.  Their relatively high median extinction of $A_V \approx 1.1$ mag implies a sizable reservoir ($\sim 10^9\,M_\odot$) of cold molecular gas, consistent with the stellar masses if the host galaxies follow the scaling relation between gas mass and stellar mass seen in star-forming galaxies.}

\item{After controlling for luminosity and redshift, type~2 AGNs, independent of either stellar mass, Eddington ratio or molecular gas mass, exhibit moderately stronger star formation activity than type~1 AGNs. This poses a severe challenge to the traditional unified model of AGNs.  Given the similar gas content of the two AGN types, their difference in SFR implies a differences in star formation efficiency.}

\item{In type~1 AGNs at $z \approx 0.3$ hosted by massive  $M_* \approx 10^{10}-10^{11.8}\,M_\odot$ galaxies,  SFR correlates strongly and linearly with AGN luminosity (BH accretion rate), with no clear dependence on stellar mass.}

\item{The star formation that connects with BH accretion rate occurs primarily on small, kpc scales.}

\item{The strong correlation observed between SFR and Eddington ratio is mostly driven by their mutual dependence on AGN luminosity. However, after removing this effect, SFR still correlates mildly with $\lambda_{\rm E}$, especially for $\lambda_{\rm E}\ga0.3$ characteristic of slim disks.}

\item{The tendency for the ${\rm SFR}-\lambda_{\rm E}$ relation to become more prominent for high $\lambda_{\rm E}$ suggests that positive feedback by AGN outflows from super-Eddington accretion may contribute to elevating circumnuclear star formation.}
\end{enumerate}

\acknowledgments
 We are grateful to an anonymous referee for constructive comments and suggestions. We thank Jinyi Shangguan for helpful discussions. This work was supported by the National Science Foundation of China (11721303, 11991052) and the National Key R\&D Program of China (2016YFA0400702).  

\vspace{5mm}
\software{Astropy \citep{2013A&A...558A..33A, 2018AJ....156..123A},  Matplotlib \citep{Hunter:2007}, Numpy \citep{numpy}, Scipy \citep{scipy} }

\end{document}